\documentclass[iop]{emulateapj}

\usepackage{graphicx,amsmath,amssymb,amsfonts}
\usepackage{color}
\usepackage[plainpages=false, colorlinks=true, anchorcolor=blue,
linkcolor=blue, citecolor=blue, bookmarks=false]{hyperref}

\shorttitle{Tidally-Driven Disk Structures} \shortauthors{Oh et al.}
\slugcomment{Accepted for publication in the ApJ}

\newcommand{\kms}{{\rm\;km\;s^{-1}}}
\newcommand{\kmskpc}{{\rm\;km\;s^{-1}\;kpc^{-1}}}
\newcommand{\kpc}{{\rm\;kpc}}
\newcommand{\pc}{{\rm\;pc}}
\newcommand{\Myr}{{\rm\;Myr}}
\newcommand{\Gyr}{{\rm\;Gyr}}
\newcommand{\Msun}{{\rm\;M_\odot}}
\newcommand{\dunit}{{\rm\;\Msun\pc^{-3}}}

\newcommand{\rtr}{r_{\rm tr}}
\newcommand{\brho}{\bar\rho}
\newcommand{\Rperi}{R_{\rm peri}}
\newcommand{\rcomp}{r_{\rm ptb}}
\newcommand{\Mcomp}{M_{\rm ptb}}
\newcommand{\Ncomp}{N_{\rm ptb}}

\newcommand{\sech}{\;\mathrm{sech}}
\newcommand{\ttime}{t_{\rm tail}}
\newcommand{\Stail}{\Sigma_{\rm tail}}
\newcommand{\itail}{i_{\rm tail}}
\newcommand{\eps}{\mathcal{E}}
\newcommand{\Ttail}{\mathcal{T}_{\mathrm{tail}}}
\newcommand{\Teff}{\mathcal{T}_{\mathrm{eff}}}
\newcommand\simgt{\lower.5ex\hbox{$\; \buildrel > \over \sim \;$}}
\newcommand\simlt{\lower.5ex\hbox{$\; \buildrel < \over \sim \;$}}
\newcommand{\Farm}{\mathcal{F}}
\newcommand{\Fmax}{\Farm_{\mathrm{max}}}
\renewcommand{\deg}{^{\circ}}

\begin{document}

\title{Physical Properties of Tidal Features of Interacting Disk Galaxies: \\
  Three-dimensional Self-consistent Models}
\author{Sang Hoon Oh$^{1,2}$, Woong-Tae Kim$^{2,3}$, and Hyung Mok Lee$^{2,3}$}
\altaffiltext{1}{Division of Computational Sciences in Mathematics,
 National Institute for Mathematical Sciences, Daejeon 305-811, Republic of Korea; email: shoh@nims.re.kr}
\altaffiltext{2}{Department of Physics and Astronomy, Seoul National
  University, Seoul 151-742, Republic of Korea}
\altaffiltext{3}{Center for Theoretical Physics (CTP), Seoul National University, Seoul 151-742, Republic of Korea}

\begin{abstract}
Using self-consistent three-dimensional (3D) $N$-body simulations, we
investigate the physical properties of non-axisymmetric features in a
disk galaxy created by a tidal interaction with its companion. The
primary galaxy consists of a stellar disk, a bugle, and a live halo,
corresponding to Milky-Way type galaxies, while the companion is
represented by a halo alone.  We vary the companion mass and the
pericenter distance to explore situations with differing tidal strength
parameterized by either the relative tidal force $P$ or the relative
imparted momentum $S$. We find that the formation of a tidal tail in
the outer parts requires $P\gtrsim 0.05$ or $S\gtrsim0.07$. A stronger
interaction results in a stronger, less wound tail that forms earlier.
Similarly, a stronger tidal forcing produces stronger, more loosely
wound spiral arms in the inner parts. The arms are approximately
logarithmic in shape, with both amplitude and pitch angle decaying with
time. The derived pattern speed decreases with radius and is close to
the $\Omega-\kappa/2$ curve at late time, with $\Omega$ and $\kappa$
denoting the angular and epicycle frequencies, respectively. This
suggests that the tidally-induced spiral arms are most likely kinematic
density waves weakly modified by self-gravity. Compared to the
razor-thin counterparts, arms in the 3D models are weaker, have a
smaller pitch angle, and wind and decay more rapidly. The 3D density
structure of the arms is well described by the concentrated and
sinusoidal models when the arms are in the nonlinear and linear
regimes, respectively. We demonstrate that dynamical friction between
interacting galaxies transfers the orbital angular momentum of one
galaxy to the spin angular momentum of the companion halo.
\end{abstract}



\section{INTRODUCTION}
\label{sec:intro}

Spiral arms play an important role in galactic evolution in diverse
ways (e.g., \citealt{but96,kor04,but13,sel14} and references therein).
They exert non-axisymmetric torque to stars and gas clouds in galaxy
rotation and cause their radial migrations, leading to secular density
changes of the disks (e.g., \citealt{foy10,ros12,bab13,kk14}). They
also provide sites for active star formation, which is either triggered
or organized by the stellar spiral potentials (e.g.,
\citealt{elm86,elm95,sle96,ber96,sei02}; see also \citealt{mck07} and
\citealt{dob14} for review).\footnote{In a different point of view,
\citet{mue76} introduced a concept of stochastic self-propagating star
formation to explain spiral structures, although it appears to produce
flocculent arms rather than grand-design arms
\citep{ger78,jun94,sle95}.} Therefore, understanding the nature and
properties of spiral arms is crucial to understand secular, chemical,
and dynamical evolution of disk galaxies.

In terms of lifetime of spiral arms, the theory of spiral structure has
forked into two branches: long-lived quasi-stationary density waves
\citep{lin64,lin66} and short-lived transient features
\citep{too69,too72}. The first picture requires that self-gravity plays
a key role in organizing density waves into a self-sustained global
pattern that rotates almost rigidly about the galaxy center
\citep{ber89a,ber89b,low94,ber96}. It successfully predicts systematic
offsets between narrow dust lanes and star-forming regions associated
with the arms (e.g., \citealt{rob69}; see also Section 6.4.3 of
\citealt{bt08}), although it does not address the origin of the density
waves. In the second picture, on the other hand, spiral arms are
transient waves that wind up over time as they propagate inward in the
radial direction. In this case, spiral arms last only for a few
rotation periods. Numerical simulations of isolated disk galaxies show
that noises inherent in a stellar disk (e.g.,
\citealt{fuj11,gra12,gra13,bab13}) or perturbations provided by giant
molecular clouds (e.g., \citealt{don13}) are amplified as they swing
from leading to trailing configurations (e.g.,
\citealt{gol65,jul66,too81}), and form transient but recurrent spiral
arms. Such swing-amplified arms in isolated galaxies are usually ragged
with multiple arms rather than being grand-design spirals with
prominent two arms.

Observations indicate that among samples that include both non-barred
and barred spiral galaxies, the probability of having grand-design arms
is higher for galaxies with companions (e.g.,
\citealt{kor79,elm82,elm87}).\footnote{Having a bar increases the
probability to possess grand-design arms especially in binary systems
\citep{elm82}.}. Recently, \citet{ken15} analyzed spiral structures in
a sample of galaxies from the \emph{Spitzer} Infrared Nearby Galaxies
Survey and found that the strength of grand-design arms is rather
tightly correlated with tidal forcing from nearby companion galaxies,
while the arm morphologies depend very weakly on the galaxy parameters
such as stellar mass, gas fraction, disk/bulge ratio, rotational
velocity, etc. This indicates that a large fraction of grand-design
arms, especially for very strong arms, are driven most likely by tidal
interactions.

\citet{too72} pioneered a numerical study for tidal interaction of disk
galaxies using noninteracting test particles, and found that extended
structures such as tidal tails and bridges develop in the course of
galaxy encounters. Subsequent $N$-body simulations focused on very
strong encounters that lead to galaxy mergers or significant
transformation of disk morphologies (e.g.,
\citealt{far82,her90b,her92,bar92,mih94,bar98,naa03,cox06,cha14}). Some
studies concentrated on the orbital parameters required to reproduce
the observed morphologies and kinematic features of interacting
galaxies (e.g., \citealt{how90,elm91,sal93,sal00a}), while others
investigated properties of tidally-induced spiral arms (e.g.,
\citealt{sun87,byr92,don94,sal00b}). In particular, \citet{sun87} found
that the spiral arms generated in a cold disk by weak tidal interaction
wind up from Sc to Sa appearances. \citet{byr92} showed that inner
spiral arms are created if the tidal strength parameter $P$ (see below
for definition) is larger than $0.01$, suggesting that even a low-mass
companion can tidally excite grand-design arms if the pericenter
distance is small enough. To characterize the tidal strength,
\citet{elm91} instead used the $S$ parameter (see also below for
definition) that takes into account the interaction duration, finding
that a tidal encounter with $S>0.019$  deforms the outer disk into an
``ocular'' shape.

Numerical simulations often show that tidally-induced arms are
transient and posses the characteristics of kinematic density waves.
For example, \citet{don94} found that spiral arms driven by tidal
forces resemble kinematic density waves when self-gravity is weak. In
modeling tidal interactions of the M51/NGC 5195 system, \citet{sal00b}
found that spiral arms in M51 cannot be described by a single pattern
speed, indicative of kinematic density waves. \citet{bab13} showed that
spiral arms in an isolated galaxy generated by swing amplification of
random perturbations have a pattern speed that decreases with the
galactocentric radius $R$ (see also \citealt{don13,mic14}).
\citet{cha14} also found that the pattern speed of $m=2$ arms driven by
a satellite on a bound eccentric orbit decreases with radius.

To quantify how the physical properties of the tidally-driven arms
depend on $S$, \citet[hereafter Paper I]{oh08} ran a series of $N$-body
simulations by considering a stellar disk inside a fixed halo
interacting with a point mass companion on a prescribed parabolic
orbit. Paper~I found that a tidal bridge consists of the disk particles
pulled out by the tidal perturbations, with their epicycle phases
locked to the companion, while a tidal tail forms as strongly perturbed
near-side particles overtake mildly perturbed far-side particles.
Paper~I also found that a stronger encounter produces stronger, more
open arms that start to develop earlier at smaller $R$. The arm pattern
speed turned out to be a decreasing function of $R$ even when the arms
are strongest, and converges to the $\Omega - \kappa/2$ curve as they
decay. This suggests that tidally-driven arms are unlikely to be
quasi-stationary density waves. Similar results were obtained by
\citet{dob10} who included a gaseous disk as well in simulating the
M51/NGC 5195 system, although gravity of the gaseous component tends to
increase the arm pattern speed. \citet{str11} suggested that these are
caustic waves maintained by coherent epicycle oscillations triggered by
tidal forcing.

While the results of the Paper~I are informative in assessing the
quantitative effects of tidal perturbations on the arm properties, they
were based on highly idealized galaxy models. First of all, Paper~I
considered an infinitesimally-thin, two-dimensional (2D) stellar disk.
This not only neglects nonplanar motions but also overestimates
self-gravity of the disk, making the spiral arms stronger than in disks
with finite thickness. In addition, Paper~I employed fixed
gravitational potentials for the halo/bulge and the perturbing
companion. This precludes the possibility of dynamical friction
occurring due to their gravitational reactions to the tidal
perturbations. Moreover, the galaxy and the companion in Paper~I were
set to follow the prescribed parabolic orbits, which ignores their
orbital decay caused by angular momentum loss. By evolving the system
in the frame in which the galaxy remains stationary, Paper~I also
ignored indirect forces arising from the orbital motion of the galaxy
relative to the center of mass of the system, potentially suppressing
the growth of lopsided spiral modes in the stellar disk (e.g.,
\citealt{ada89,ost92}).

To overcome the caveats mentioned above, we in this paper extend
Paper~I by considering a self-consistent three-dimensional (3D) galaxy
model in which the stellar disk has a finite thickness and a central
bugle and a dark halo are represented by live particles rather than by
fixed potentials. The perturbing companion is modeled by a live halo
alone for simplicity. As in Paper~I, we control the strength of a tidal
encounter by varying two parameters: the pericenter distance and the
galaxy-to-companion mass ratio. Our objective is four-fold. First, we
wish to measure the physical properties of tidally-induced spiral arms
in more realistic 3D models and compare them with those in the
razor-thin counterpart. We will also find the quantitative dependence
of the arm properties on both $S$ and $P$. Second, \cite{spr99} and
\cite{dub99} showed that the formation of tidal tails depends on the
shape of the galactic potential. We will show that the tail formation
depend not only on the potential shape but also on the tidal stregnth.
Third, we want to explore 3D density structures inside spiral arms and
compare them with the analytic formulae proposed by \citet{cox02}.
Finally, we will show that dynamical friction of the companion in the
course of a tidal interaction can be a source of the spin angular
momentum of the primary halo.

This paper is organized as follows. In section~\ref{sec:method_model},
we describe our galaxy model and introduce the parameters for tidal
strength. In Section~\ref{sec:properties}, we present the properties of
tidal tails formed in the outer parts and spiral arms in the inner
parts. In Section~\ref{sec:armdensity}, we analyze the 3D density
structures of spiral arm in the radial and vertical directions and
compare them with the prediction of \cite{cox02}. In
Section~\ref{sec:orbit_decay}, we explore the orbital decay of the
galaxy caused by dynamical friction, which results in spin-up of the
initially non-rotating halo. Finally, we summarize our results and
discuss their astronomical implications in Section~\ref{sec:summary}.

\begin{figure}
\hspace{0.25cm}\includegraphics[angle=0, width=8cm]{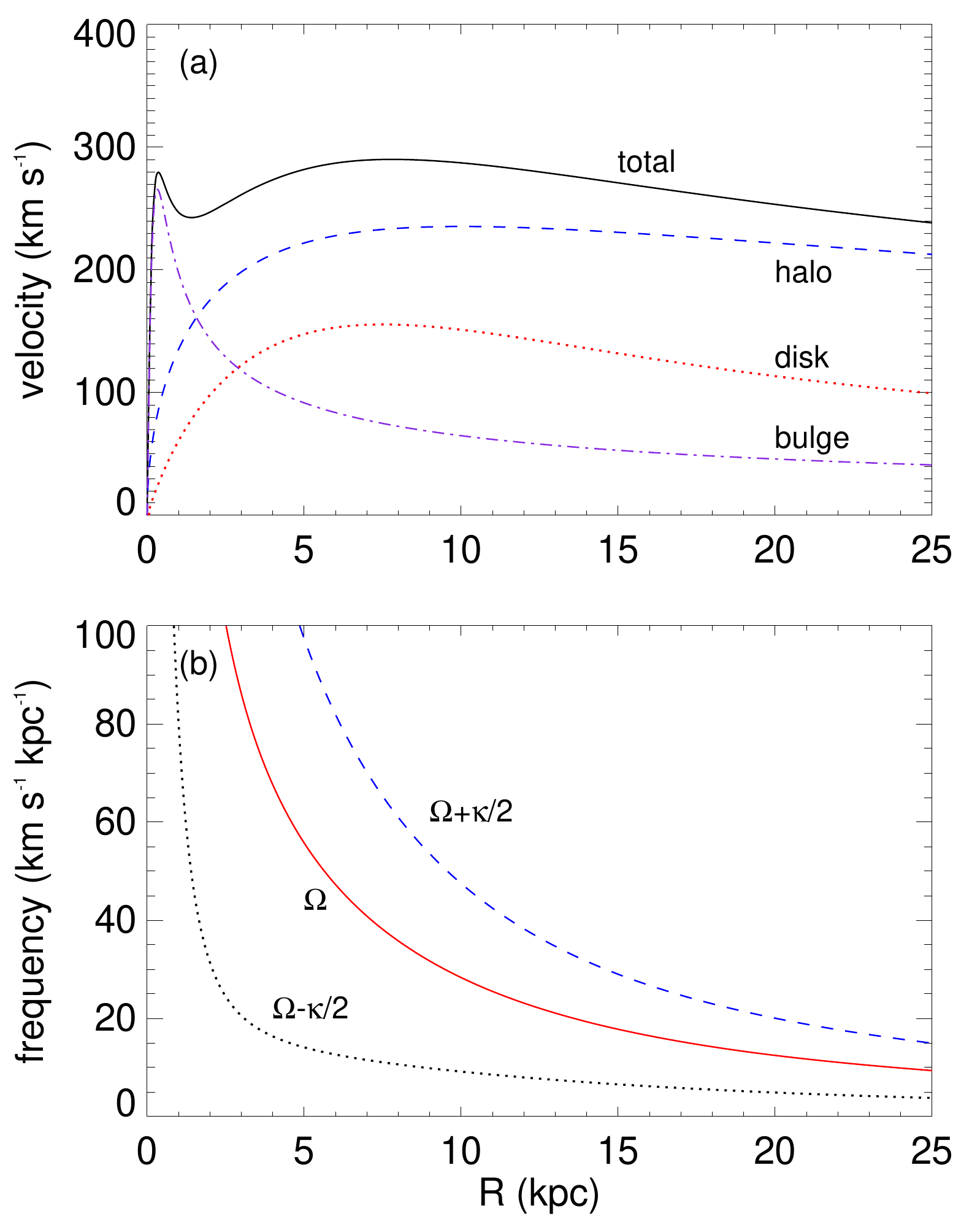}
\caption{\label{fig:rotprofile}
  (a) Radial profile of the circular velocity of the initial disk
and the contribution of each component. (b) Radial distributions of the
characteristic angular frequencies $\Omega$ (solid), $\Omega-\kappa/2$
(dotted), and $\Omega+\kappa/2$ (dashed).}
\end{figure}

\section{MODEL}\label{sec:method_model}

\subsection{Galaxy Models}

In this paper we use 3D $N$-body simulations to study the tidal
interaction of a disk galaxy with its companion. The galaxy consists of
a stellar disk, a spherical bulge, and a dark matter halo; the effect
of a gaseous component is not considered in the present study.   For
the halo, we adopt a truncated \citet{her90a} model
\begin{equation}\label{eq:rho_halo}
  \rho_h(r) =
  \left\{
    \begin{array}{l@{\quad\textrm{for}\, r\;}l@{\;\rtr,}}
      \dfrac{M_h}{2\pi} \dfrac{r_h }{r(r+r_h)^{3}}, & \leq \\
       0, & >
    \end{array}
  \right.
\end{equation}
with the mass $M_h=5.15 \times 10^{11} \Msun$, the scale radius
$r_h=10\kpc$, and the truncation radius $\rtr=200\kpc$, within which
$90\%$ of $M_h$ is enclosed.  For the bulge, we take a Plummer
sphere
\begin{equation}\label{eq:rho_bulge}
  \rho_b(r)  = \frac{3 M_b}{4\pi} \frac{r_b^2 }{(r^2+r_b^2)^{5/2}},
\end{equation}
with $M_b = 9.80 \times 10^{9} \Msun$ and $r_b=0.23\kpc$.

The stellar disk is initially axisymmetric with density distribution
\begin{equation}
  \label{eq:rho_disk}
  \rho_d(R,z) = \frac{M_d}{4 \pi h_0 R_d^2} \exp\left(-\frac{R}{R_d}\right)
  \sech^2\left(\frac{z}{h_0}\right),
\end{equation}
where $M_d=5.17 \times 10^{10} \Msun$ is the total disk mass, $R_d =
3.4\kpc$ is the radial scale length, and $h_0=0.33\kpc$ is the vertical
scale height.  The corresponding stellar surface density is $\Sigma_d =
37.4 \Msun\rm\;kpc^{-2}$ at $R=10\kpc$. The mass ratio of each
component of the galaxy is  $0.9 M_h : M_b : M_d = 9.1 : 0.2 : 1$, with
the total mass $M_{\rm gal}=5.30\times 10^{11}\Msun$ within the
truncation radius.

\begin{deluxetable*}{lcccccccc}
\tabletypesize{\scriptsize} \vspace{-0.5cm} \tablecaption{Model
parameters and simulation results\label{tab:model}}%
 \tablewidth{0pt}
\tablehead{
 \colhead{\begin{tabular}{l@{}} Model\tablenotemark{a}\\ (1) \end{tabular}} &
 \colhead{\begin{tabular}{c@{}} $\Mcomp / M_g$    \\ (2) \end{tabular}} &
 \colhead{\begin{tabular}{c@{}} $\Rperi$ (kpc)     \\  (3) \end{tabular}} &
 \colhead{\begin{tabular}{c@{}} $P$ \\ (4) \end{tabular}} &
 \colhead{\begin{tabular}{c@{}} $S$ \\ (5) \end{tabular}} &
 \colhead{\begin{tabular}{c@{}} $t_{\rm tail}$ \\ (6) \end{tabular}} &
 \colhead{\begin{tabular}{c@{}} $\tan i_{\rm tail}$ \\ (7)  \end{tabular}} &
 \colhead{\begin{tabular}{c@{}} $\Sigma_{\rm tail}/\Sigma_{20}\!$\ \\ (8) \end{tabular}} &
 \colhead{\begin{tabular}{c@{}} $\Fmax$ \\ (9) \end{tabular}}
}
 \startdata
 TA1 & 0.44 & 25 & 0.410 & 0.248 &  0.12   &  0.69   &  30.9  & 0.173   \\
 TA2 & 0.44 & 35 & 0.180 & 0.166 &  0.16   &  0.61   &  23.1  & 0.108   \\
 TA2H& 0.44 & 35 & 0.175 & 0.163 &  0.15   &  0.61   &  25.5  & 0.113   \\
 TA3 & 0.44 & 45 & 0.096 & 0.121 &  0.21   &  0.55   &  13.6  & 0.066   \\
 TB1 & 0.22 & 25 & 0.202 & 0.134 &  0.19   &  0.45   &  14.2  & 0.111   \\
 TB2 & 0.22 & 35 & 0.089 & 0.091 &  0.25   &  0.45   &  10.2  & 0.061   \\
 TB3 & 0.22 & 45 & 0.046 & 0.065 &  0.35   &  0.39   &   7.6  & 0.039   
\enddata
 \tablenotetext{a}{Model TA2H is identical to model TA2 except that
the former employs 20 times more particles than the latter. }
\end{deluxetable*}

Figure \ref{fig:rotprofile} plots the radial distribution of the
circular velocity and the contribution of each component as well as the
characteristic angular frequencies $\Omega$ and $\Omega\pm\kappa/2$ in
the initial disk. The presence of a strong bulge makes the
$\Omega-\kappa/2$ curve rise steeply toward the center, which
suppresses the formation of a bar in our tidal encounter models (e.g.,
\citealt{too81,sel00}). To find the equilibrium velocity distribution
of the disk particles under the total gravitational potential, we solve
the axisymmetric Jeans equations using the method described in
\cite{her93} and \cite{spr05}. We ensure that the dark matter halo and
bulge do not rotate initially. For the disk, the radial profile of the
radial velocity dispersion is chosen such that the Toomre $Q$ stability
parameter of the 3D disk is similar to that of the 2D counterpart
considered in Paper~I; we set $Q \approx 2$ over $5 \kpc\lesssim R
\lesssim 15\kpc$, which is large enough to suppress a spontaneous
growth of spiral structures when evolved in isolation.

We model the perturbing companion using a dark matter halo alone, which
follows the truncated Hernquist profile
\begin{equation}\label{eq:rho_comp}
  \rho_{\rm ptb}(r) =      \frac{\Mcomp}{2\pi} \frac{\rcomp
  }{r(r+\rcomp)^{3}}, \;\;\text{for}\; r \leq 100\kpc,
\end{equation}
with the scale radius of $\rcomp = 5\kpc$, and $\rho_{\rm ptb}=0$ for
$r>100\kpc$. We vary the companion mass $\Mcomp$ to study the
dependence of tidal features on the strength of tidal force (see
Section \ref{sec:param}). Compared with the point-mass model considered
in Paper~I, the current extended-halo model realizes a more realistic,
smooth variation of the perturbing gravitational potential especially
when the companion is close to the pericenter.

Most of the simulations presented in this paper use the galaxy model
constructed by distributing $N_h = 4 \times 10^{5}$, $N_b = 1 \times
10^4$, and $N_d = 1 \times 10^5$ particles for the halo, bulge, and
disk of the primary galaxy, respectively, and $\Ncomp = 1 \times 10^5$
particles for the companion. We evolve our 3D galaxy model in isolation
for 2 Gyr, finding that the disk remains almost axisymmetic, with very
weak ($\sim1\%$ density variations) nonaxisymmetric features forming at
its outskirts.  This suggests that our initial disk is globally stable,
due to a strong bulge as well as quite large value $Q$ $(\sim 2)$, in
the absence of tidal perturbations. To check the dependency of
simulation outcomes on the particle numbers, we also run a
high-resolution model using $N_h = 8 \times 10^{6}$, $N_b = 2 \times
10^5$, $N_d = 2 \times 10^6$, and $\Ncomp = 2 \times 10^6$ particles.
We find that resolution does not make significant differences in the
properties of tidal features: the spiral arms in the low-resolution
model are $\sim10\%$ weaker and have a $\sim10\%$ smaller pitch angle
than in the high-resolution counterpart.

\begin{figure*}
\vspace{-0.5cm}
\hspace{0.5cm}\includegraphics[angle=0, width=18cm]{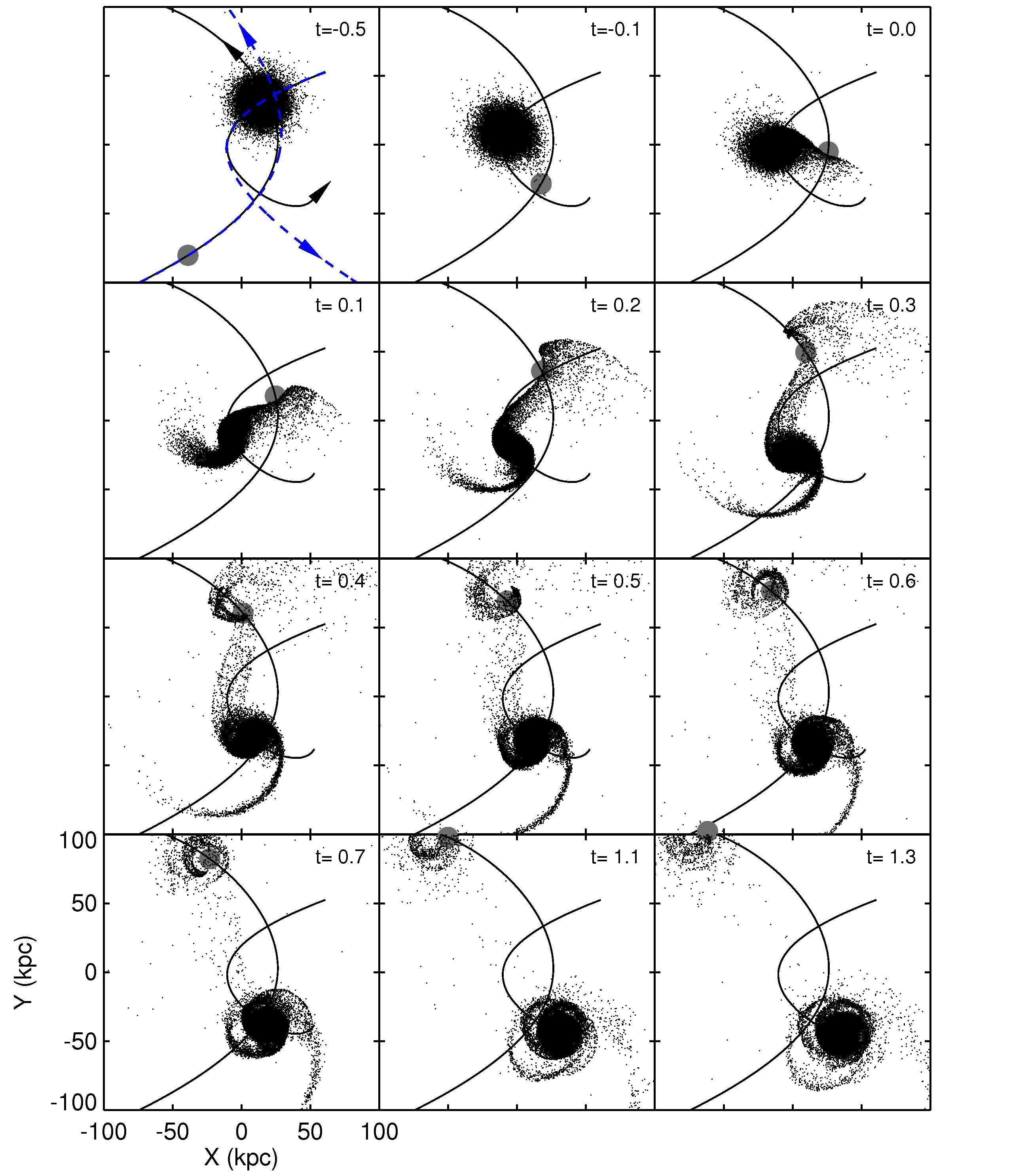}
  \caption{\label{fig:snapshots}
Distributions of the disk particles of the primary galaxy in model TA2
projected on the $X$--$Y$ plane. In each panel, the position of the
companion is marked by a grey circle, with its size indicating the
half-mass radius. The solid lines draw the actual trajectories of the
primary and companion, while the dashed lines in the top-left panel
give their Keplerian orbits that assume no change in the mass
distributions of the two galaxies.}
\end{figure*}

\subsection{Model Parameters}\label{sec:param}

Initially, we place the two galaxies at large separation, and make them
move on mutual parabolic orbits with a pericenter distance $\Rperi$
that is calculated under the assumption that the total mass of each
galaxy is concentrated at its own center of mass. The galaxy orbits are
prograde and located at the same plane as the disk of the primary
galaxy. To study tidal interactions with differing strength, we
consider six models that differ in $\Mcomp$ and $\Rperi$.

The responses of the stellar disk to the tidal field of the companion
can be parameterized by either
\begin{equation}
  \label{eq:P}
  P = \left(\frac{\Mcomp}{M_{g}}\right)
    \left(\frac{R_{g}}{\rcomp + \Rperi}\right)^3,
\end{equation}
or
\begin{equation}
  \label{eq:S}
  S = \left(\frac{\Mcomp}{M_{g}}\right)
    \left(\frac{R_{g}}{\rcomp + \Rperi}\right)^3
    \left(\frac{\Delta T}{T}\right),
\end{equation}
where $R_g$, taken equal to $7R_d$ in this work, is the characteristic
size of the primary galaxy (e.g., \citealt{spr99}), $M_g$ is the galaxy
mass within $R_g$, $\Delta T$ is time elapsed for the companion to
orbit one radian near the pericenter, and $T$ is time taken by the disk
particles at $R=R_g$ to rotate one radian about the galaxy center. The
$P$ parameter in Equation \eqref{eq:P} directly measures the tidal
force exerted by the companion relative to the gravitational force of
the primary at $R_g$ (e.g., \citealp{byr92}). On the other hand, the
$S$ parameter in Equation \eqref{eq:S} takes allowance for the duration
of tidal interaction and thus corresponds to the ratio of the linear
momentum imparted to a disk particle by the companion with respect to
its original linear momentum of galaxy rotation (e.g.,
\citealp{elm91}). Note that the current definition of $S$ differs
slightly from that given in Paper~I, since the companion now has an
extended density distribution rather than being treated as a point
mass. When the tidal perturbations are applied \emph{impulsively} over
the time scale of $\Delta T$, $S$ is a good measure of tidal strength
(e.g., Section 8.2 of \citealt{bt08}). One can thus expect that $S$ is
a better measure of tidal forcing for tidal tails formed in the outer
disk where the rotation period is longer than $\Delta T$.

Table ~\ref{tab:model} lists the parameters of each model and some
simulation outcomes. Column (1) labels each run. We choose model TA2
with $\Mcomp/M_g=0.44$ and $\Rperi=35\kpc$ as our fiducial model, which
is the 3D counterpart of model A2 of Paper~I. Model TA2H is identical
to model TA2 except that the former employs 20 times more particles
than the latter. Column (2) gives the companion mass relative to the
primary galaxy. Column (3) lists the pericenter distance of two
galaxies under the Keplerian orbits. Columns (4) and (5) give the
dimensionless interaction strengths $P$ and $S$, respectively. Columns
(6), (7), and (8) list the time $\ttime$ when the tidal tail develops
strongest, and its pitch angle $\tan i_{\mathrm{tail}}$ and surface
density $\Sigma_{\mathrm{tail}}/\Sigma_{20}$ at $t=\ttime$,
respectively. Here, $\Sigma_{20}$ is the surface density of the initial
disk at $R=20\kpc$. Finally, Column (9) gives the peak strength of the
spiral arms that are induced.

We take $10^{10}\Msun$, $1\kpc$, and $1\kms$ as the units of mass,
length, and velocity, respectively. In these units, the gravitational
constant is $G=4.289 \times 10^4$, and the corresponding unit of time
is $t_0= 0.98\Gyr$. The simulation time is set such that $t=0$
corresponds to the closest approach of the galaxies under the
prescribed parabolic orbits, but the tidal deformation of the galaxies
makes the actual time of the closest approach delayed to
$t=0.02$--0.04, with a smaller value corresponding to weaker tidal
forcing. All the simulations start from $t=-1.5$, when the two galaxies
are separated more than $280\kpc$. For model TA2, the initial positions
and velocities of the galaxies are $\mathbf{r}_1=(R_1,\theta_1, z_1) =
(80.3, -2.4, 0)$ and $\dot{\mathbf{r}}_1 = (-39.0, 0.2, 0)$ for the
primary, and $\mathbf{r}_2 =(206.3, 0.7, 0)$ and
$\dot{\mathbf{r}}_2=(-100.2, 0.2, 0)$ for the companion.

The simulations are performed using the GADGET code that employs the
Barnes-Hut hierarchical tree algorithm \citep{bar86} to solve the
Poisson equation \citep{spr01}. We adopt the force error tolerances of
$\alpha = 0.02$ and $\theta = 0.8$. The gravitational softening
parameter in the low-resolution models is set to $0.1\kpc$ for the
disk, the halo, and the companion, and $0.07\kpc$ for the bulge,
corresponding to 2.8 times the mean particle distance in each system
(e.g.~\citealt{spr01}).

\section{Properties of Tidal Features}\label{sec:properties}

Tidal perturbations disturb not only the otherwise axisymmetric disk,
producing a tidal tail and a bridge in the outer parts ($R \gtrsim 20
\kpc$) and spiral waves in the inner parts ($ 4 \lesssim R \lesssim
15\kpc$), but also the primary halo and companion galaxy.  In this
section, we explore the correlations between the physical properties of
the tidal features and the interaction strength parameters $P$ and $S$.
The deformation of the halo and companion will be presented in Section
\ref{sec:orbit_decay}.

\subsection{Tidal Tail and Bridge}\label{sec:tail}

Figure~\ref{fig:snapshots} shows morphological changes of the stellar
disk of the primary galaxy in our fiducial model TA2 projected on to
the orbital plane. The grey circle in each frame indicates the center
of mass of the companion at a given epoch, with its size corresponding
to the region encompassing a half of the companion mass. The solid
curves represent the actual trajectories of the two galaxies. The
dashed lines in the $t=-0.5$ frame draw the parabolic orbits they would
follow if their extended mass distributions were undisturbed over the
course of the encounter. The difference between the solid and dashed
orbits are due to the angular momentum loss caused by dynamical
friction, as will be discussed in Section \ref{sec:orbit_decay} in
detail.

\begin{figure*}
\hspace{0.5cm}\vspace{0cm}\includegraphics[angle=0, width=18cm]{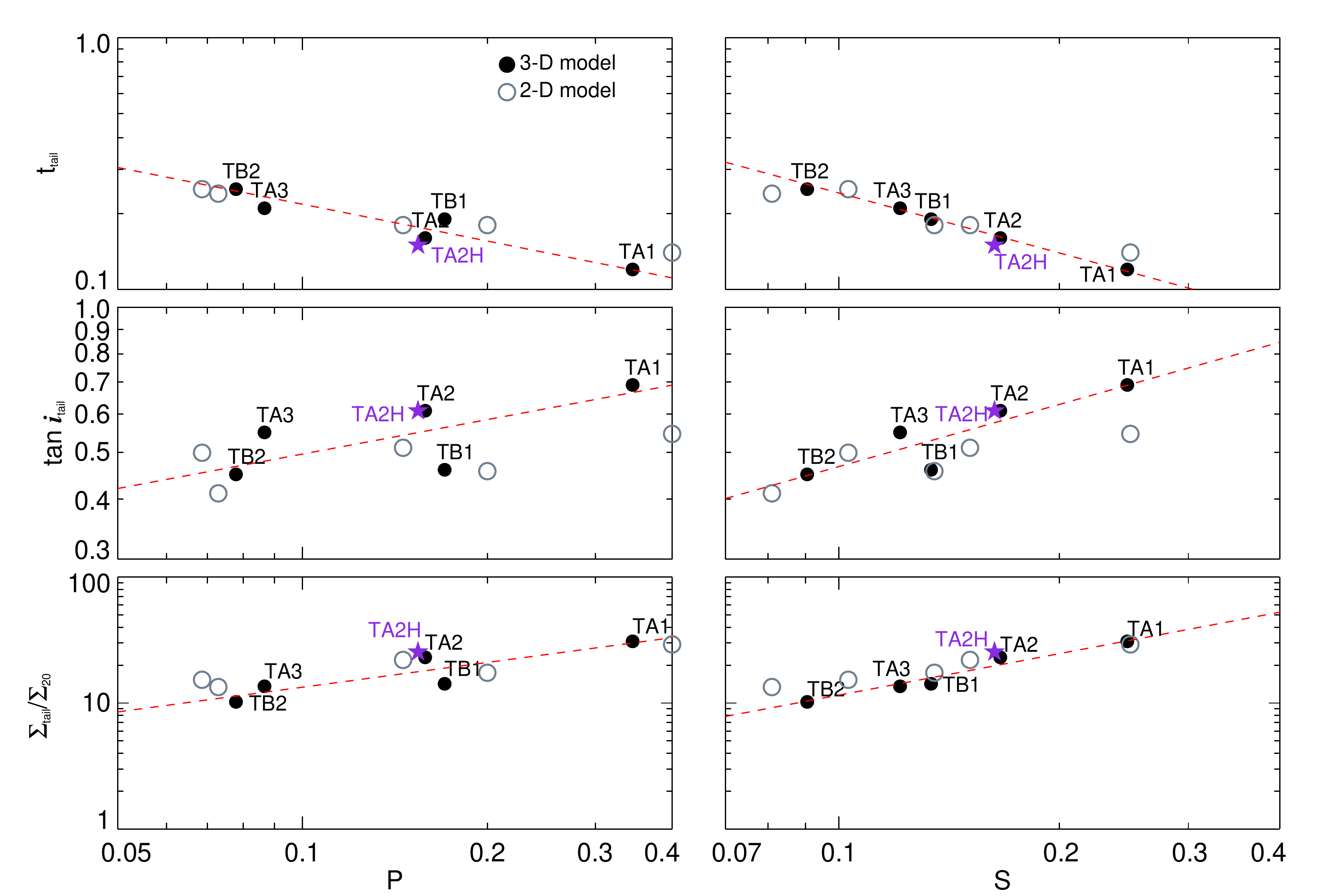}
  \caption{\label{fig:sp-tail}
  Dependence of the formation epoch $\ttime $, the
  pith angle $\itail$, and the surface density
  $\Stail$ of tidal tails on the tidal strength parameters $P$ (left) and $S$ (right).
  Filled symbols are the current 3D results, while the 2D razor-thin results are compared as open circles.
  The dashed lines are the best fits: $\ttime = 0.07 P^{-0.49}$, $\tan \itail = 0.1
 P^{0.24}$, and $\Stail/\Sigma_{20} = 60 P^{0.65}$ in terms of $P$;
 $\ttime = 0.04 S^{-0.79}$, $\tan \itail = 1.25 S^{0.43}$, and
 $\Stail/\Sigma_{20} = 144 S^{1.1}$ in terms of $S$.
}
\end{figure*}

At early time when the galaxies are separated widely ($t \lesssim
-0.1$), tidal perturbations are so weak that the orbits are parabolic
and the disk of the primary remains almost axisymmetric.  As the
perturbing galaxy approaches the pericenter, the tidal force begins to
excite the epicycle motions of stars in the outer disk, first shaping
them into a bridge at the near side to the companion ($t\sim0-0.2$) and
then a tail at the opposite side ($t\sim0.1-0.3$).  In the bridge, the
epicycle phases of the constituting particles are forced to be aligned
to the companion in such a manner that the radial velocities and the
gradient of the azimuthal velocities are always maximized along the
line connecting the galaxies (Paper~I). On the other hand, the tidal
tail develops as strongly perturbed particles at the near side catch up
and overlap with mildly perturbed particles at the far side (Paper~I;
\citealt{pfl63, too72}). The tail is approximately logarithmic in
shape, indicating that its pitch angle is independent of $R$.

The tails and bridges produced in our models are transient and decay
after $t\sim0.2-0.4$.  As the companion recedes, the tidal force
becomes weaker and is thus unable to keep the coherency of the forced
epicycle phases of the bridge particles. Composed of particles gathered
from a wide range of radii in the unperturbed disk, the tails have a
large velocity dispersion, tending to disperse with time. They
disappear almost completely at $t\sim 1.0$. A small fraction ($\sim
2\%$) of the disk particles that achieve velocities larger than the
escape velocity become unbound,  either captured by the companion or
escaping from the system. With no further perturbations from the
companion at $t>1.0$, on the other hand, the bound particles once in
the bridge or tail spread out, following galaxy rotation with large
eccentricities. The disk becomes increasingly featureless over time.

Paper~I found that the formation time, shape, and strength of the tails
in 2D models are well correlated with $S$. To study how the
correlations change in the current self-consistent 3D models, we define
the tail formation epoch, $\ttime$, as the time when it becomes densest
at $R=20\kpc$, and measure its pitch angle $\itail$ and surface density
$\Stail$ at that time. Figure~\ref{fig:sp-tail} plot these values as
filled symbols against $P$ and $S$; these are also tabulated in Columns
(6)--(8) of Table ~\ref{tab:model}. With weak tidal interaction, model
TB3 does not produce a readily identifiable tail, suggesting that the
formation of tidal tail requires $P\gtrsim 0.05$ or $S \gtrsim 0.07$.
This threshold is probably a lower limit since all of our models
consider in-plane, prograde encounters, the most favorable condition
for the tail development. A stronger tail tends to form earlier and has
a larger pitch angle. The dashed lines are the best power-law fits to
the 3D results:  $\ttime = 0.07 P^{-0.49}$, $\tan \itail = 0.1
P^{0.24}$, and $\Stail/\Sigma_{20} = 60 P^{0.65}$ in terms of $P$;
$\ttime = 0.04 S^{-0.79}$, $\tan \itail = 1.25 S^{0.43}$, and
$\Stail/\Sigma_{20} = 144 S^{1.1}$ in terms of $S$. As expected, the
tail properties are better correlated with $S$ than $P$: the linear
fitting coefficients are $\sim0.82$ and $\sim0.93$ against $P$ and $S$,
respectively. For comparison, we also plot the 2D results taken from
Paper~I as as open circles. Overall, the 2D and 3D results agree very
well, although the 3D tails are slightly weaker, which is probably
caused by weaker disk gravity due to finite disk thickness.
Figure~\ref{fig:sp-tail} shows that the tail in model TA2H with high
resolution has almost the same properties as in model TA2, suggesting
that our results do not depend sensitively on resolution.

\begin{figure}
\hspace{0.2cm}\includegraphics[angle=0, width=8cm]{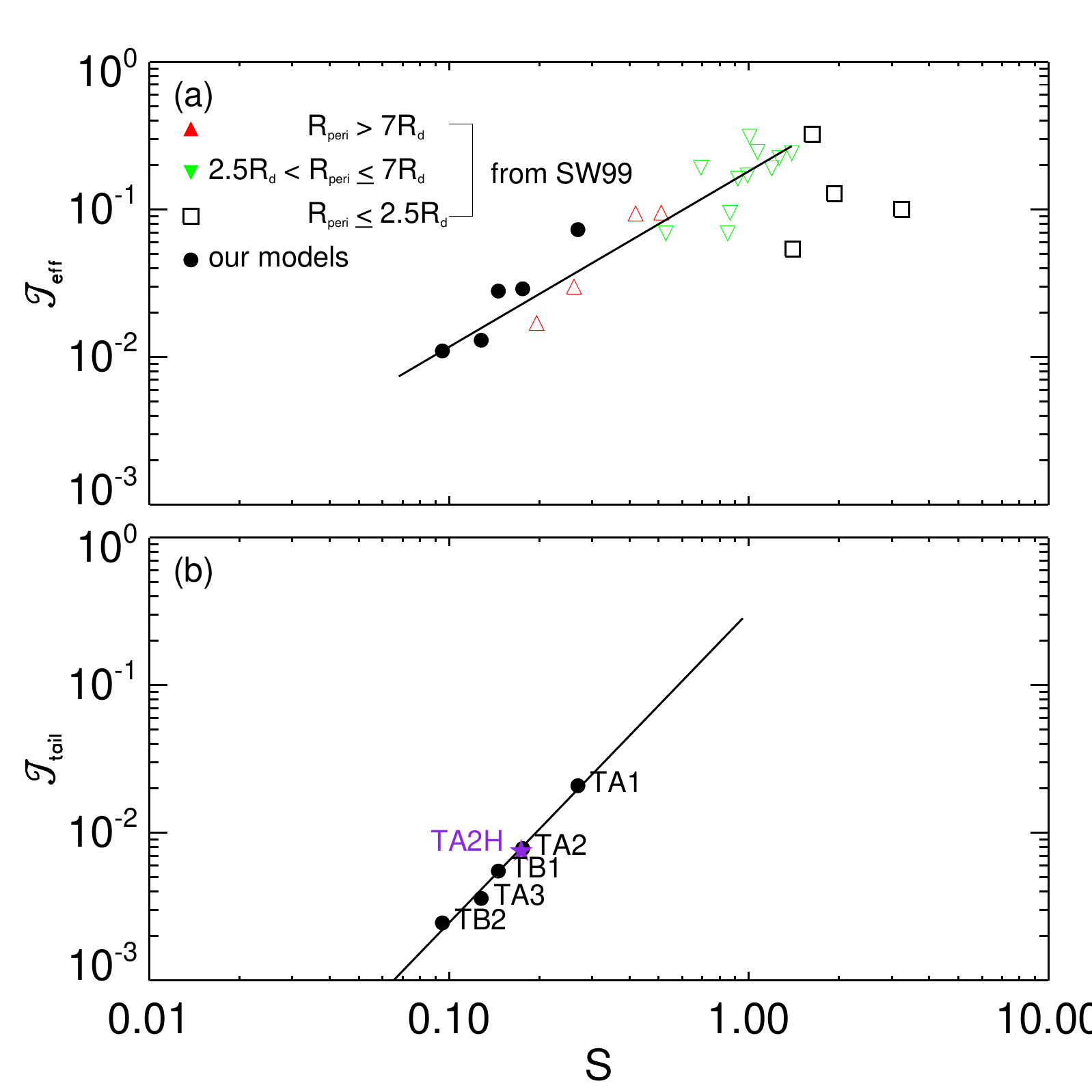}
  \caption{\label{fig:spteff}
   Dependence on $S$ of (a) $\Teff$, the maximum fraction of the disk particles at $R>10R_d$, and
   (b) $\Ttail$, the fraction of the disk particles in the tail at
   $R>10R_d$. Filled circles are from our 3D simulations.
   The results of \citet{spr99} for models with $R_{\mathrm{peri}}/R_d>7$ (regular triangles),
   $2.5<R_{\mathrm{peri}}/R_d \leq 7$ (inverted triangles), and
   $R_{\mathrm{peri}}/R_d \leq 2.5$ (squares) are compared.
   The solid lines are the best power-law fits:
    $\Teff = 0.16 S^{1.1}$ and
    $\Ttail = 0.31 S^{2.1}$.}
\end{figure}

\begin{figure*}
\hspace{0.5cm}\includegraphics[angle=0, width=17cm]{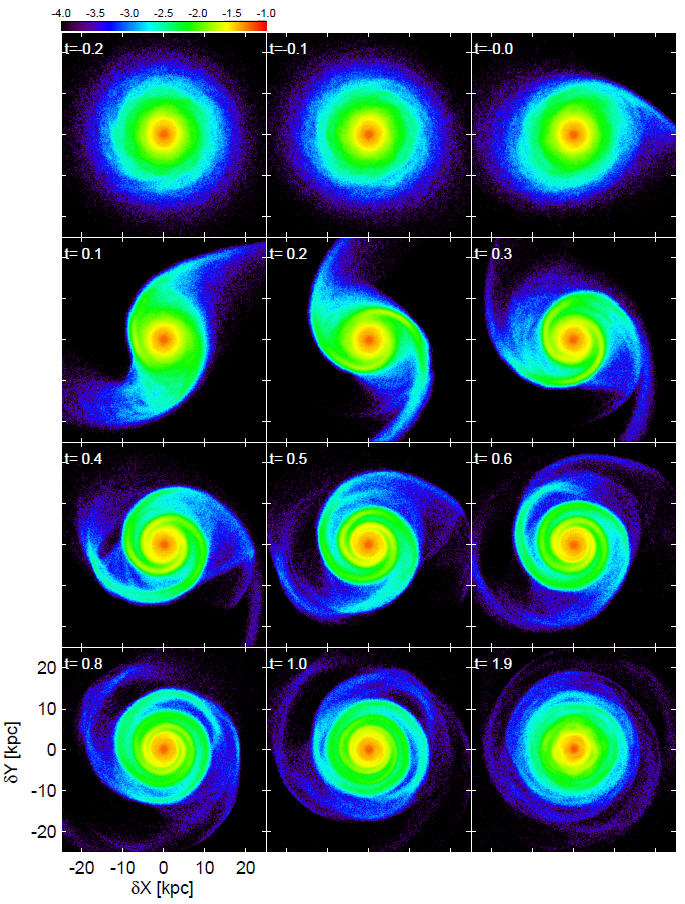}
  \caption{\label{fig:surf_xy}
   Zoom-in snapshots of the disk surface density in model TA2H at the times shown in
   Figure \ref{fig:snapshots}. The coordinates are shifted so as to make the
   center of mass of the disk located at the origin ($\delta X=\delta Y=0$).  The colorbar labels
   $\log(\Sigma/10^4 \Msun \mathrm{pc}^{-2})$. The spiral arms become strongest at $t\approx0.3-0.4$ and
   decay afterwards.}
\end{figure*}

Another way to quantify the tail strength is to use the tidal response
introduced by \citet{spr99} who ran a number of $N$-body simulations
for galaxy mergers with differing halo mass and interaction strength.
They defined $\Teff$ as the maximum fraction of the disk particles in
each model that reach distances beyond $10R_d$ ($=34\kpc$ in our
models) from the disk center in the course of interaction.
Figure~\ref{fig:spteff}(a) plots $\Teff$ from our tail-forming models
(filled circles) as well as the models (open symbols) from
\citet{spr99} as a function of $S$: regular triangles are for models
with $R_{\mathrm{peri}}/R_d
> 7$; inverted triangles for $2.5<R_{\mathrm{peri}}/R_d \leq 7$;
squares for $R_{\mathrm{peri}}/R_d \leq 2.5$. All of our models have
$R_{\mathrm{peri}}/R_d > 7$. Despite differences in the galaxy models
such as potential depth and rotation curve, etc., our results are
overall in good agreement with the extrapolation of \citet{spr99} for
$S<1$. Note that in very strong encounter models with $S>1$  in
\citet{spr99} represented by open squares, the companion directly
passes through the inner parts of the primary disk, for which the $S$
parameter relying on a distant tide approximation is not useful in
characterizing the interaction strength. Except for these very
strong-encounter models, there is a good correlation between $\Teff$
and $S$ with some scatters. The best fit for models with
$R_{\mathrm{peri}}/R_d > 2.5$ is $\Teff = 0.16 S^{1.1}$, which is shown
as the solid line in Figure~\ref{fig:spteff}(a).

\begin{figure*}[!t]
\hspace{0.5cm}\includegraphics[angle=0,width=17cm]{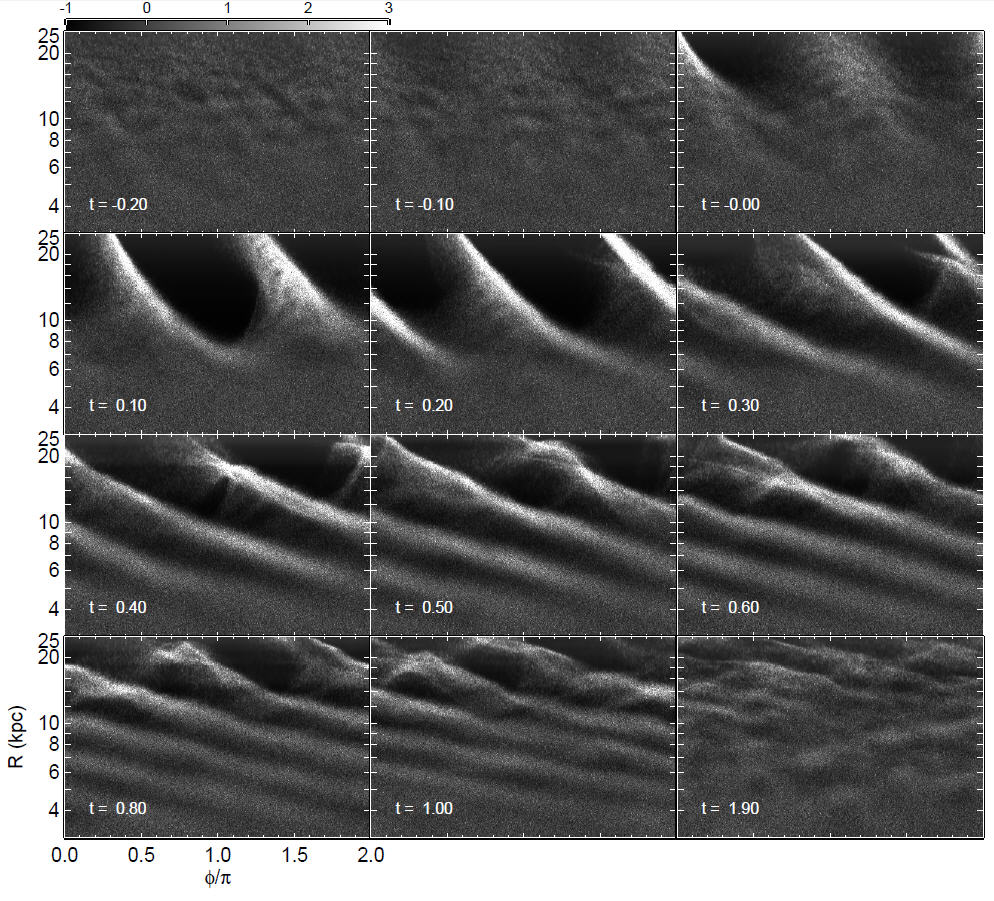}
\caption{\label{fig:surf_phir}
 Distributions of the perturbed surface density
    $\delta \Sigma/\Sigma_0$ of model TA2H in the $\phi$--$\ln R$ plane.
    When $t\lesssim0$, the disk is almost featureless.
    When $0\lesssim t\lesssim 0.2$, $\delta \Sigma$ is dominated by a bridge and a tail
    at $R \gtrsim 15\kpc$, which become weak after $t\sim 0.3$.
    The spiral arms at $R \lesssim 15\kpc$ are approximately
    logarithmic and wind progressively with time. The greyscale bar labels $\log \delta \Sigma/\Sigma_0$.}
\end{figure*}

Since $\Teff$ includes the particles not only in a tail but also
escaped or captured to the companion, to measure the tail strength we
similarly define $\Ttail$ as the fraction of the disk particles
consisting only of a tail at $R>10R_d$. Figure~\ref{fig:spteff}(b)
plots $\Ttail$ for our models as a function of $S$. The solid line
draws the best fit $\Ttail = 0.31 S^{2.1}$, which has a tighter
correlation than the $\Teff$--$S$ relation. The $\Ttail$--$S$
relationship can be explained as follows. Let us assume that the tidal
perturbations are applied impulsively near the pericenter during the
time interval $\Delta T$. The radial velocity increment at $R_g$ is
then given by $\delta v_R=2GM_pR_g\Delta T/(\Rperi+r_p)^3$. This
enhances the epicycle amplitudes of the disk particles to $\delta R
\approx \delta v_R/\kappa_g$. With the epicycle frequency $\kappa_g
=\sqrt{2}\Omega_g = (2GM_g/R_g^2)^{1/2}$ at $R_g$ for flat rotation, it
follows that $\delta R/R_g = \sqrt{2} S$. Since the perturbed mass that
goes into the tail is proportional to $\Sigma_0\delta R^2$, one obtains
$\Ttail \propto S^2$, similar to our numerical results.

\subsection{Spiral Arms}\label{sec:sdw}

We have seen in the previous section that the tidal force from the
companion provides strong perturbations to epicycle orbits of particles
in the outer disk, resulting in a tidal tail and a bridge. Tidal
disturbances in the inner part of the disk are not as strong as in the
outer parts, but can nevertheless induce spiral arms there (e.g.,
\citealt{too69,don91}). Figure~\ref{fig:surf_xy} plots close-up
snapshots of the stellar surface density in logarithmic scale in model
TA2H projected onto the orbital plane, with the center of mass of the
disk shifted to the origin ($\delta X=\delta Y=0$). At $t=0.1$, the
outer edges of the arms are quite sharp, which is a common feature of
tidally-induced arms (e.g., \citealt{str90,elm91,elm11,dob10}).

To delineates the spiral structure, Figure~\ref{fig:surf_phir} plots
the perturbed surface density, $\delta \Sigma \equiv \Sigma -
\Sigma_0$, relative to the initial surface density $\Sigma_0$ in the
$\phi$--$\ln R$ plane. Before the companion passes by the pericenter
($t \lesssim -0.1$), the disk appears almost axisymmetric, without
forming a bar or spiral structures, indicating again that the disk is
globally stable in isolation. Tidal disturbances excite epicycle
motions of the particles in the inner disk, which are coherently
organized to build up a well-defined two-armed global spiral pattern.
With $Q\sim2$, the effect of swing amplification on the growth of the
spiral arms is not significant (Paper~I). It is apparent that the arms
are approximately logarithmic. In model TA2H, they become strongest at
$t=0.2-0.4$ and decay as they keep winding out over time. In these
single encounter experiments, no spiral arms are apparent after $t=1.5$
(see also Fig.~\ref{fig:surf_xy}).

\begin{figure}
\hspace{0.2cm}\includegraphics[angle=0, width=8cm]{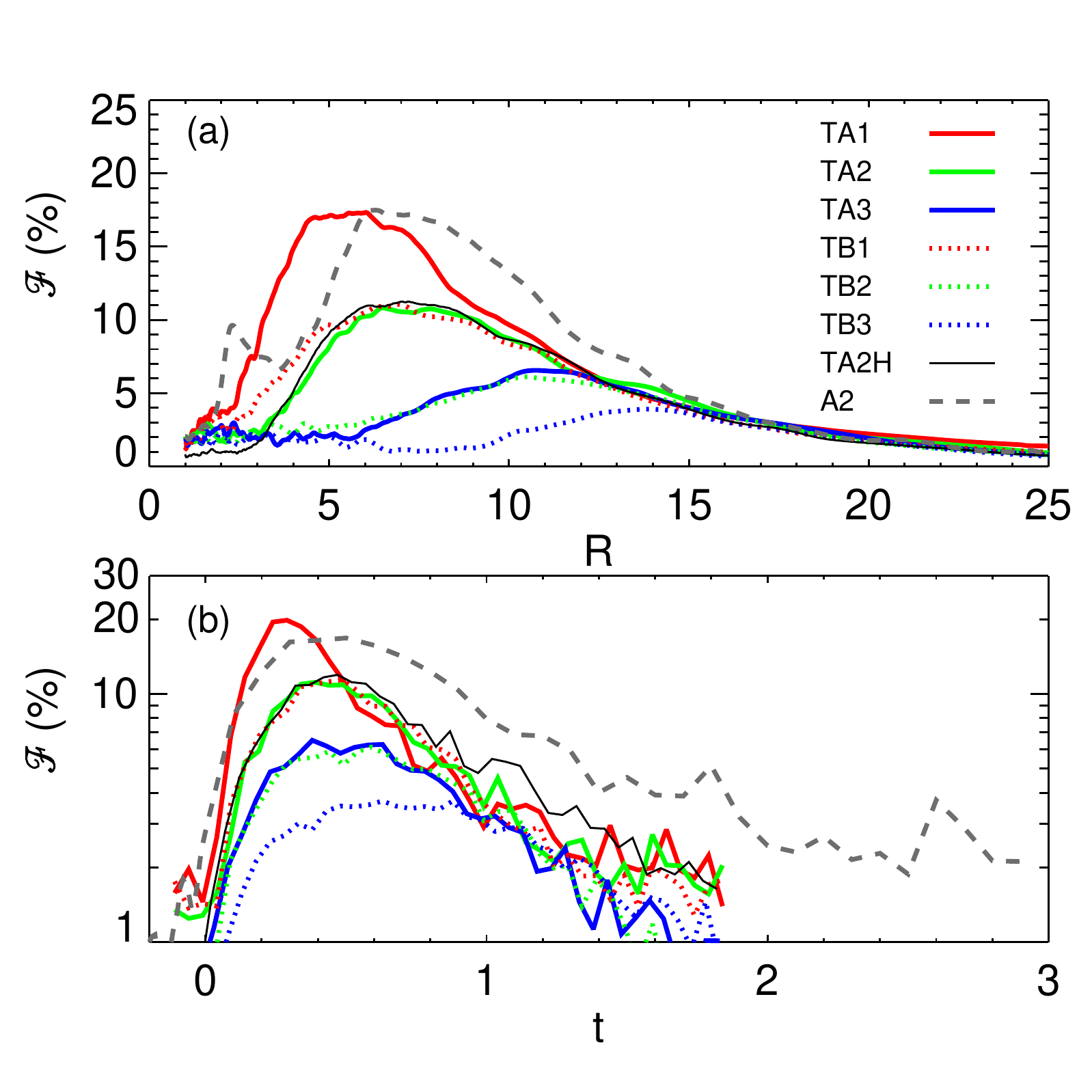}
  \caption{\label{fig:F-R}
    (a) Radial dependence of the arm strength $\Farm$ averaged over the time interval
    $\Delta t =0.45$ centered at the time of the peak strength. (b)
    Temporal dependence of $\Farm$ averaged over $5\kpc\leq R \leq
    10\kpc$ for models TA1, TA2, and TB1, and over $8\kpc\leq R \leq
    15\kpc$ for models TA3, TB2, and TB3. After the peak, $\Farm$
    decays as $\sim\exp(-t/0.5\Gyr)$. In both panels, the results of the 2D model
    A2 from Paper~I are compared as dashed lines.}
\end{figure}

\subsubsection{Arm Strength}

One of the most important parameters that control the responses of gas
flows across a stellar spiral arm is the arm strength (e.g.,
\citealt{kk14,kkk14}). It has often been customary to quantify the arm
strength using the dimensionless parameter
\begin{equation}\label{eq:F}
  \Farm \equiv \frac{2 \pi G \tilde \Sigma_{m=2}}{R \Omega^2},
\end{equation}
where $\tilde\Sigma_{m=2}$ denotes the amplitude of the $m=2$ Fourier
mode of $\Sigma$. Note that $\Farm$ effectively measures the maximum
gravitational force due to the spiral arms as a fraction of the
axisymmetric gravitational force in the unperturbed disk
\citep{rob69,shu73,kim02,kim06,she06}.\footnote{For $m$-armed
tightly-wound spirals with pitch angle $i (\ll1)$ and radial wavenumber
$k_R=m/(R\tan i)$, the corresponding gravitational potential at $z=0$
of a razor-thin disk is $\Phi_{m} = \tilde \Phi_{m} \exp(im\phi + ik_R
R)$ with $\tilde \Phi_{m} = -2\pi G \tilde \Sigma_{m}/ k_R$ (e.g., Eq.\
(6.31) of \citealt{bt08}). Therefore, $\Farm \equiv |d\Phi_{m}/dR|_{\rm
max}/(R\Omega^2) = m|\tilde \Phi_{m}|/(\tan i R^2\Omega^2) = 2\pi G
\tilde \Sigma_m /(R\Omega^2)$.}

\begin{figure*}
\hspace{0.5cm}\includegraphics[angle=0, width=16cm]{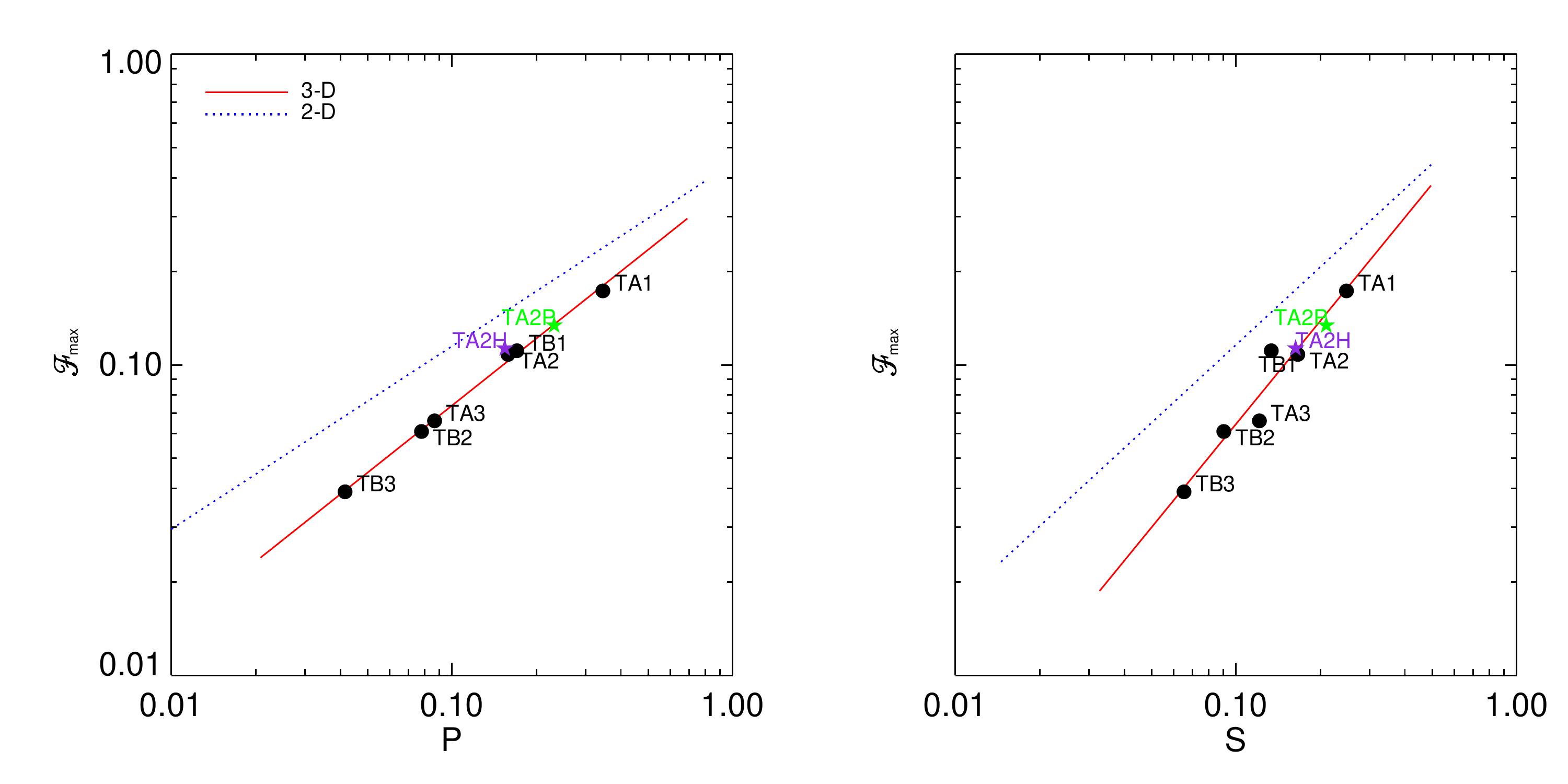}
  \caption{\label{fig:sp-f}
    Dependence of the peak arm strength $\Fmax$ on (a) $P$ and (b) $S$.  A stronger tidal forcing
    results in stronger arms.  The solid lines are the best fits,
     $\Fmax = 0.39 P^{0.7}$ and $\Fmax = 0.82 S^{1.1}$, to the 3D
     results, while the dotted lines draw the 2D results adopted from Paper~I.}
\end{figure*}

Figure~\ref{fig:F-R}(a) plots the radial variation of $\Farm$ averaged
over the time interval $\Delta t = 0.45$ centered at the epoch when the
arms attain the maximum strength in each model, while
Figure~\ref{fig:F-R}(b) gives the temporal change of $\Farm$ averaged
over a range of radii where the arms are strong. Definitely, a weaker
tidal interaction produces weaker arms that grow more slowly at larger
galactocentric radii: they are maximized at $R \sim 5 - 10\kpc$ for the
intermediate encounter models TA1, TA2, and TB1 with $P>0.17$ (or $S>
0.13$), and at $R\sim 8$--$15\kpc$ for the weak encounter models TA3,
TB2, and TB3 with $P<0.1$ (or $S<0.12$). The difference between the
results of models TA2 and TA2H is insignificant, indicating again that
resolution does not affect the simulation outcomes much. For
comparison, Figure~\ref{fig:F-R} also plots the results of the 2D model
A2 taken from Paper~I as dashed lines. Note that the arms in model A2
have $\Farm$ about 70\% stronger than in the 3D counterpart, which is
primarily due to the overestimated self-gravity in the 2D, razor-thin
geometry.

\begin{figure}
\hspace{0.5cm}\includegraphics[angle=0, width=8cm]{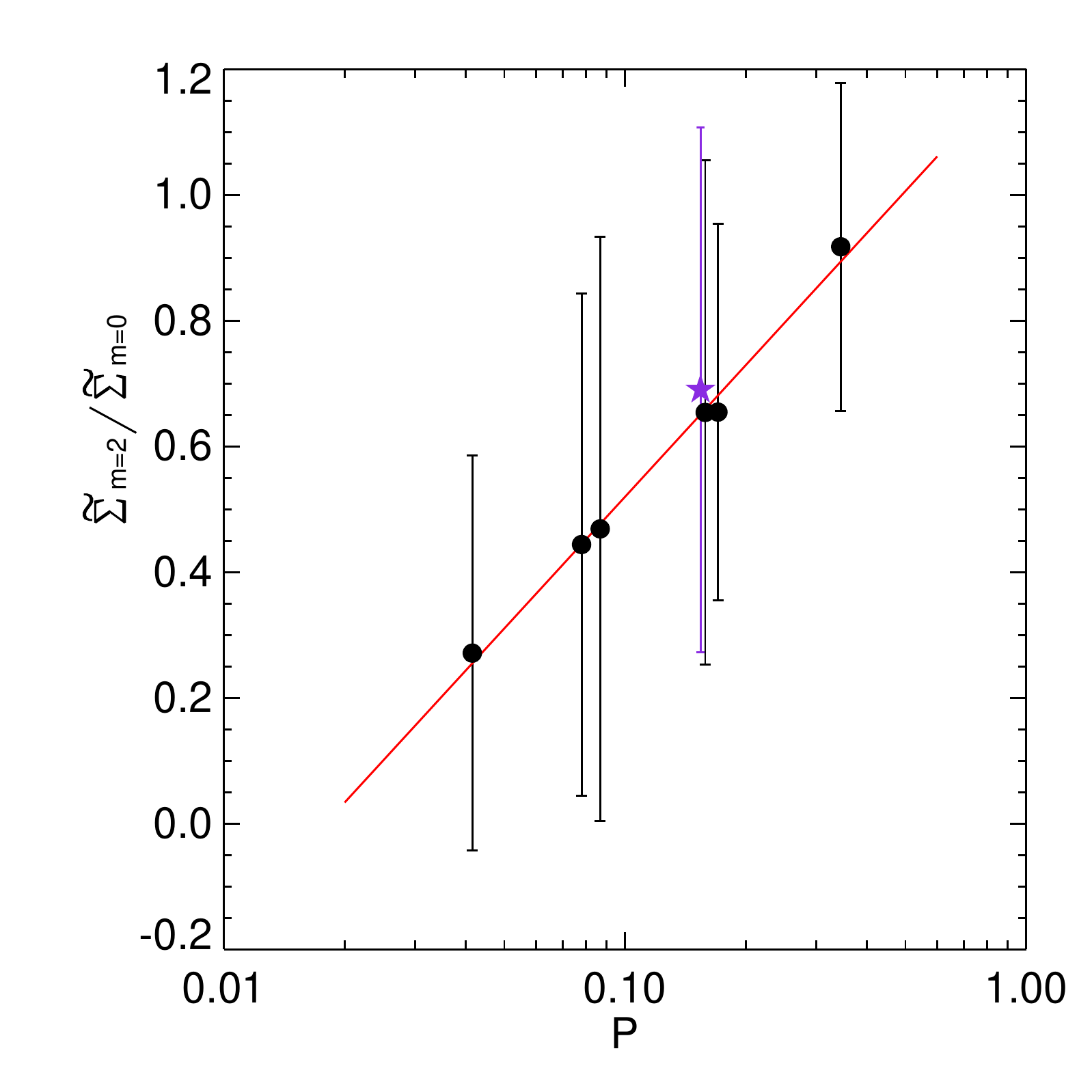}
   \caption{\label{fig:m2m0_P}
     Dependence of the relative amplitude
     $\mathcal{R}= \tilde\Sigma_{m=2}/\tilde\Sigma_{m=0}$ of the arms
     on the $P$ parameter. The solid circles mark the averaged over
     $0\leq t \leq 1$ and $5\kpc \leq R \leq 15\kpc$, with the errorbars indicating
     the standard deviations. The solid line is the best fit: $\mathcal{R} = 1.22 + 0.70 \log P$.}
\end{figure}

Figure~\ref{fig:sp-f} plots the peak arm strength $\Fmax$ as functions
of $P$ and $S$. The solid lines represent the best power-law fits
$\Fmax=0.39P^{0.7}$ and $\Fmax=0.82S^{1.1}$ to the 3D results, with the
linear correlation coefficients of $0.99$ and $0.95$ relative to $P$
and $S$, respectively.  These can be compared to the 2D results of
$\Fmax=0.79 S^{0.83}$ from Paper~I shown as dotted lines. Observations
usually measure the arm strength in terms of the relative Fourier
amplitude $\mathcal{R} \equiv \tilde\Sigma_{m=2}/\tilde\Sigma_{m=0}$
(e.g., \citealt{rix93,pat01,ken15}). In order to facilitate comparisons
with observations, Figure~\ref{fig:m2m0_P} presents the dependence on
$P$ of $\mathcal{R}$ calculated from our 3D results. The filled circles
plot the averages over $0\leq t\leq 1$ and $5\kpc \leq R\leq 15\kpc$,
with the errorbars corresponding to the standard deviations. The solid
line is the best fit: $\mathcal{R} = 1.22 + 0.70 \log P$.

The spiral arms achieve the full strength at $t\approx0.3$--$0.6$ with
smaller values corresponding to stronger tidal interaction, after which
they begin to decay. This is mainly caused by the increase in the
velocity dispersions of the disk particles. First of all, the growth of
spiral arms requires the particles at different radii to gather into
the arms, which increases the velocity dispersions (Paper~I). Also,
heating of the disk particles due to gravitational scattering off the
arms becomes efficient once the arms are strong enough, inhibiting
further growth of the arms (e.g., \citealt{sel84,bin01}). The increased
velocity dispersions make the epicycle orbits in the arms kinematically
less coherent. Figure~\ref{fig:F-R}(b) shows that the amplitudes of the
arms in the current 3D models decrease with time almost exponentially,
with a characteristic timescale of $\sim 0.5$\;Gyr, about twice faster
than in the 2D models of Paper~I. The faster decay of the arms in the
3D models is presumably due to weaker disk gravity in a stratified
disk, which makes the arms wind out at a faster rate as well.

\begin{figure}
\hspace{0.2cm}\includegraphics[angle=0, width=8.5cm]{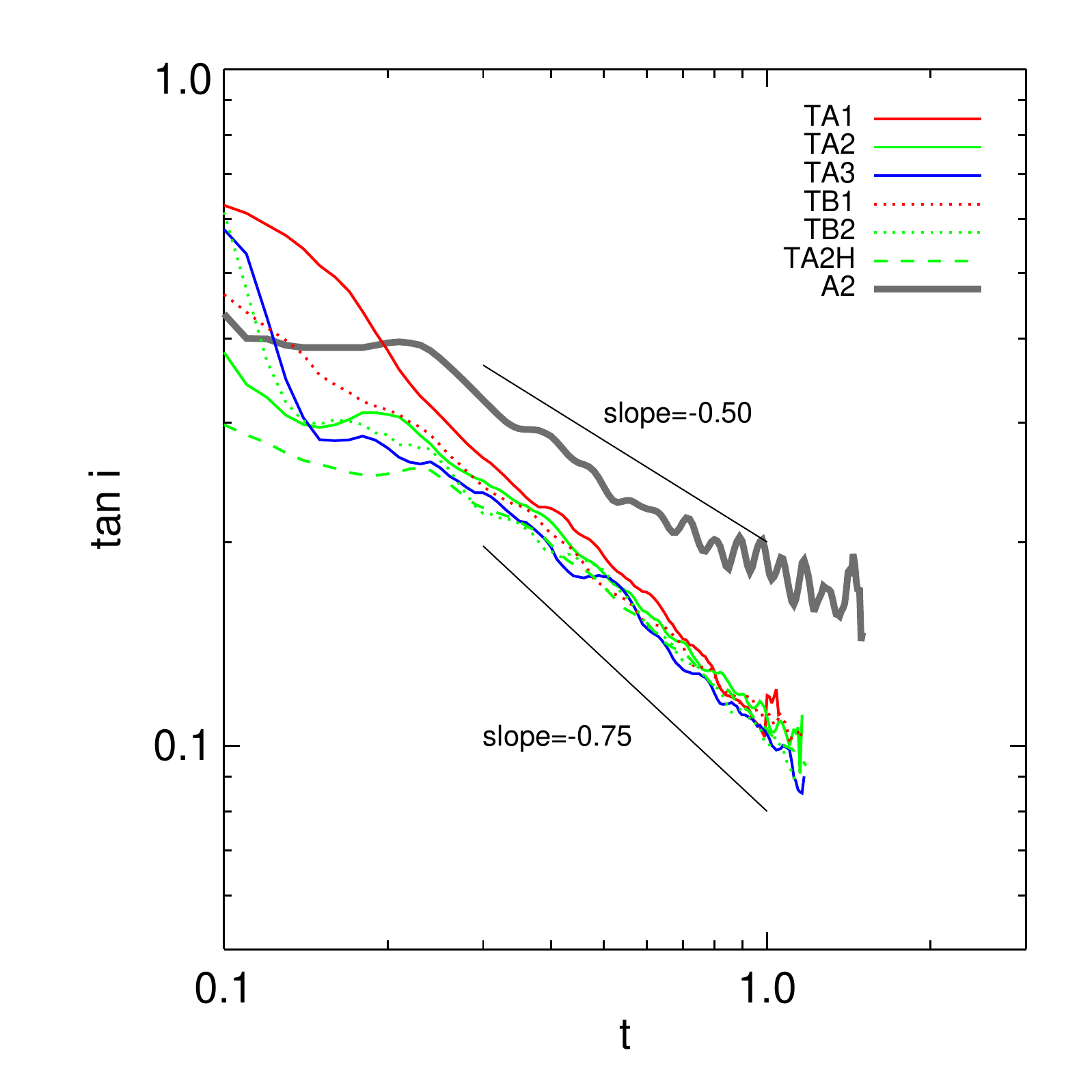}
  \caption{\label{fig:winding}
  Temporal variations of the pitch angle $i$ of the spiral arms located
  at $5\kpc\leq R \leq 10\kpc$ for the intermediate forcing models TA1, TA2, and TB1, and at
  $8\kpc\leq R \leq  15\kpc$ for the weak forcing models TA3, TB2, and TB3. The result of the 2D,
  razor-thin model A2 is compared as a thick grey line.
  The arm pitch angle in the 3D models decays as $t^{-0.75}$ after the peak, which is steeper than
  $t^{-0.6}$ in the 2D model.}
\end{figure}

\subsubsection{Pitch Angle}

Figure~\ref{fig:surf_phir} shows that the tidally-induced spiral arms
are approximately logarithmic in shape over a wide range of radii. This
appears consistent with the observational results that grand design
spirals are close to be logarithmic in optical and near-infrared images
(e.g., \citealt{gro98,sei06,dav12,mar12}). As in Paper~I, we calculate
the arm pitch angle at a given time by first finding the Fourier
coefficients in $\phi$ and $\ln R$ defined as
\begin{equation}  \label{eq:Fourier}
A(p) = \frac{1}{N_p}\sum_{j=1}^{N_p} \exp [i(2\phi_j + p\ln R_j)],
\end{equation}
where $N_p$ is the number of particles located at $R=5$--$10\kpc$ for
the intermediate tidal-forcing models with $P>0.17$ (or $S>0.13$), and
at $R=8$--$15\kpc$ for the weak forcing models with $P<0.1$ (or
$S<0.12$), $(R_j, \phi_j)$ are the coordinates of the $j$-th particle,
and $p$ is a real number corresponding to the slope of a two-armed
spiral in the $(\ln R, \phi)$ plane.  We then calculate $p_{\rm max}$
that maximizes $|A(p)|$ and calculate the pitch angle of the arms
through $\tan i = 2/p_{\rm max}$ at each time (e.g.,
\citealt{sel84,sel86}).

Figure~\ref{fig:winding} plots the temporal changes of the arm pitch
angles for all 3D models. The 2D result of model A2 from Paper~I is
also compared. When the arms stand out initially, they have moderate
pitch angles amounting to $i\sim 15\deg-20\deg$. After attaining the
peak amplitude, the arms in the 3D models wind out over time as $\tan
i\propto t^{-0.75}$, which is steeper than $\tan i \propto t^{-0.6}$ in
the 2D models. The faster winding rates in 3D models are again because
of reduced self-gravity in vertically-extended disks. Note that the
pitch angle of purely kinematic density waves with no self-gravity
decays as $\tan i \propto t^{-1}$ (e.g., Section 6.2 of
\citealt{bt08}).

\begin{figure*}
\hspace{0.5cm}\includegraphics[angle=0, width=17cm]{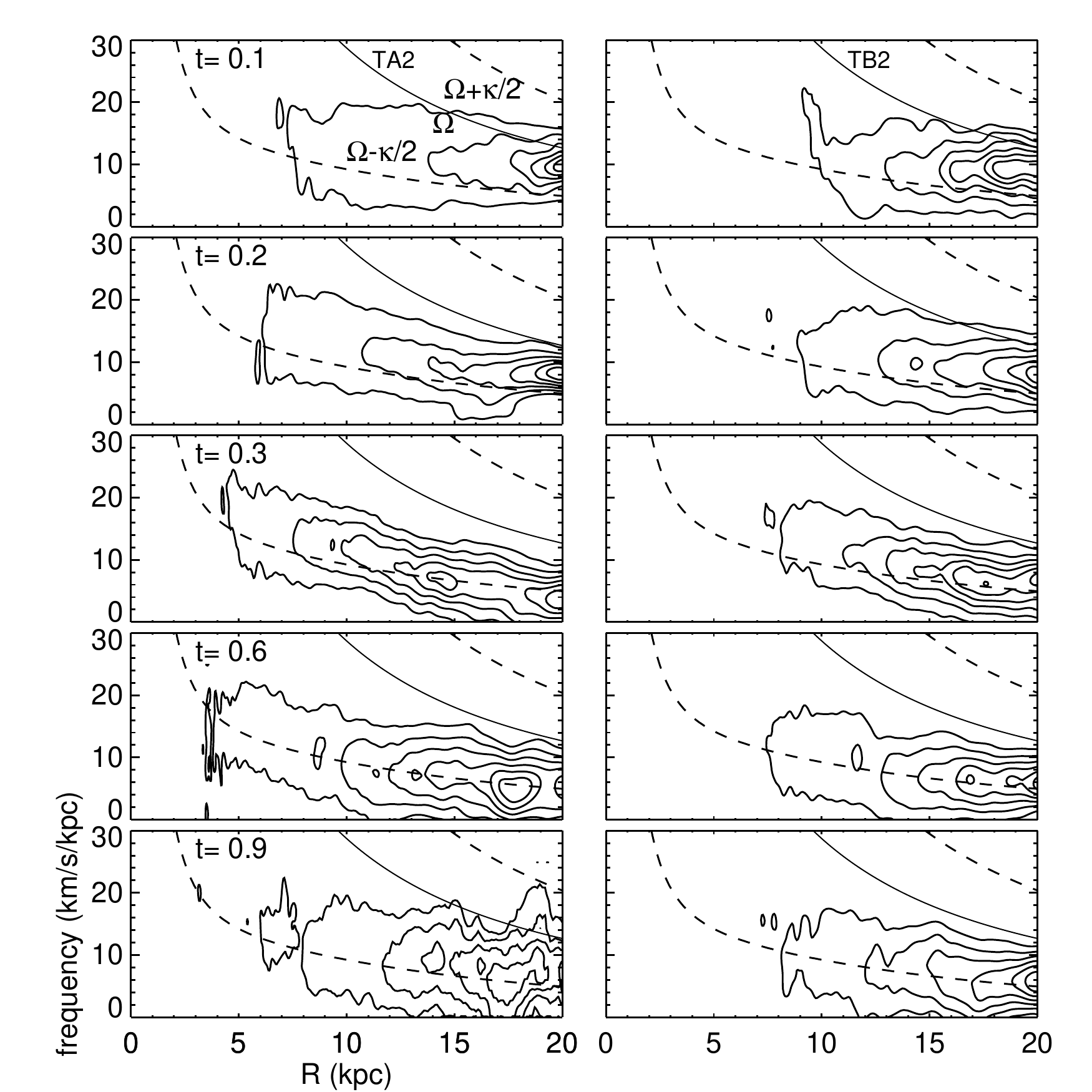}
 \caption{\label{fig:omegap}
Contours of the cross-correlation $C(R, \varphi, t)$ of the perturbed
surface density in the radius-frequency plane for models TA2 (left) and
TB2 (right) at $t=0.1$, 0.2, 0.3, 0.6, and 0.9. The solid and dashed
lines draw $\Omega$ and $\Omega\pm\kappa/2$ curves. At $t\lesssim 0.2$,
the cross-correlation is dominated by the tidal bridge and tail at
$R\gtrsim 17\kpc$, while the spiral arms dominate at $5\kpc\lesssim R
\lesssim 15\kpc$ for $t\gtrsim 0.3$. Note that $\Omega_p$ of the arms
is very close to the $\Omega-\kappa/2$ curve.}
\end{figure*}

\begin{figure*}
\hspace{0.5cm}\includegraphics[angle=0, width=17cm]{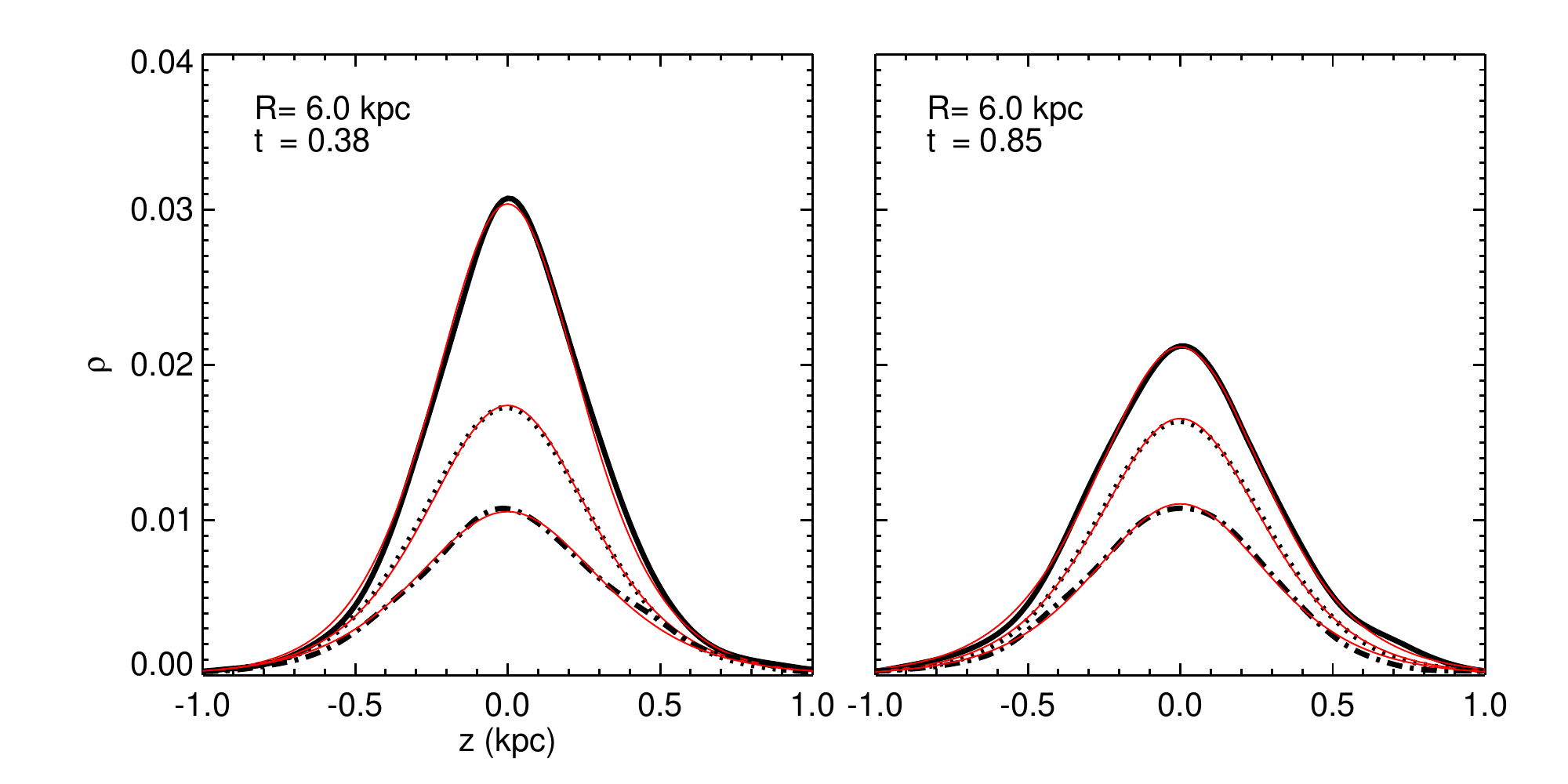}
  \caption{\label{fig:rhoz_arm}
  Vertical distributions of the stellar surface density in the arm
  regions of model TA2H at $R=6\kpc$ for (a) $t=0.38$
  when the arms are strong, and (b) $t=0.85$ when the arms become weak.
  The black solid, dotted, dot-dashed lines draw the total density
  $\rho$, the azimuthally-averaged density $\brho$, and the
  perturbed density $\rho_1=\rho-\brho$, respectively. The thin red
  lines are fits based on the $\mathrm{sech}^2(z/h)$ function, with the
  scale heights
  $h=0.33$, $\bar{h}=0.36$, $h_1=0.30\kpc$ at $t=0.38$  and  $h=0.37$,
  $\bar{h}=0.36$, $h_1=0.41\kpc$ at $t=0.85$, for the total,
  azimuthally-averaged, perturbed densities, respectively.}
\end{figure*}

\begin{figure*}
\hspace{0.5cm}\includegraphics[angle=0, width=17cm]{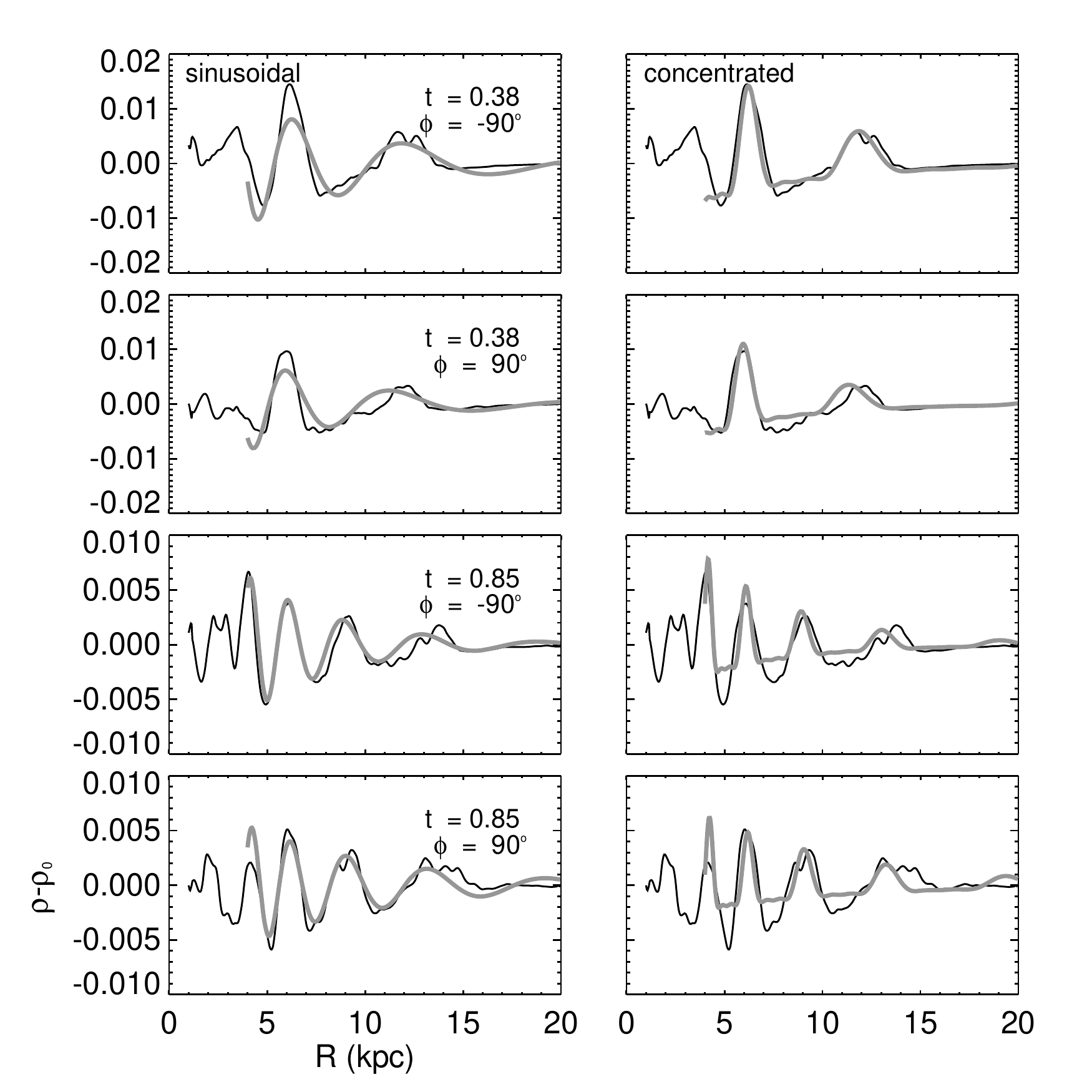}
  \caption{\label{fig:rhor_arm}
    Perturbed density $\rho_1(R,\phi,z)$ at the $z=0$ plane
    as a function of $R$ for two different azimuthal
    directions ($\phi = \pm90^{\circ}$) of model TA2H at $t=0.38$ when the
    spiral arms are quite strong (top and second rows) and at $t=0.85$
    when the arms are weak (third and bottom rows).
    The thick grey lines represent the best fits using
    Equation \eqref{eq:rho_arm} based on the sinusoidal-arm (left column)
    and the concentrated-arm (right column).}
\end{figure*}

\subsubsection{Pattern Speed}

The pattern speed of spiral arms is also an important parameter,
although it is not well constrained observationally. To calculate the
arm pattern speed induced in our models, we use the normalized
cross-correlation of the perturbed surface densities at two different
epoches separated by $\Delta t$ as
\begin{equation}\label{eq:cross}
C(R, \varphi, t)  = \frac{1}{\Sigma_0(R)^2} \int_0^{2\pi} \delta
\Sigma(R,\phi,t) \delta \Sigma(R,\phi + \varphi ,t+ \Delta t) d\phi.
\end{equation}
By taking a sufficiently small value of $\Delta t=0.1$, we find
$\varphi_{\mathrm{max}}$ that maximizes $C(R, \varphi, t)$ at given
radius and time. The instantaneous pattern speed is then given by
$\Omega_p(R,t) = \varphi_{\mathrm{max}} / \Delta t$.

Figure~\ref{fig:omegap} plots as contours the amplitudes of the
$C(R,\varphi,t)$ in the $R$--$(\varphi/\Delta t)$ plane for some
selected epoches of models TA2 (left) and TB2 (right).  The angular
frequencies $\Omega$ and $\Omega \pm \kappa/2$ from the initial
rotation curve are overplotted as a solid line and two dotted lines,
respectively. The locus of $\varphi_{\mathrm{max}}$ in each panel draws
the arm pattern speed as a function of $R$. Shortly after the
pericenter passage of the companion ($t\sim 0.1$), the perturbed
surface density is dominated by the tidal bridge rather than the arms.
The bridge has an almost fixed pattern speed at $\Omega_p \approx
10$\,km s$^{-1}$ kpc$^{-1}$, which corresponds to the angular speed of
the companion near the pericenter. At this time, the particles inside
the bridge are tidally locked to the companion. As the companion moves
away from the primary, the bridge becomes weaker and $C(R,\varphi,t)$
becomes progressively dominated by the spiral arms. At $t\sim 0.3$ when
the arms have substantial amplitudes, $\Omega_p$ is larger than $\Omega
- \kappa/2$, but only slightly. Since $\Omega_p$ is not constant over
radius, the arms produced in our models are not exactly a pattern in a
strict sense. They are rather similar to gravity-modified kinematic
density waves. Although self-gravity tends to increase $\Omega_p$, its
effect is not significant in our 3D models. At late time ($t \gtrsim
0.6$), $\Omega_p$ converges to the $\Omega - \kappa/2$ curve.

To summarize this section, spiral arms induced by a tidal interaction
in our models are not quasi-stationary density waves envisaged by
\citet{lin64}. They are rather kinematic density waves slightly
modified by gravity. Due to reduced self-gravity, spiral arms in the 3D
models are weaker and decay faster than those in the razor-thin
counterparts. The arms in vertically-stratified disks have a smaller
pitch angle and wind up more rapidly than those in the razor-thin, 2D
disks.

\section{Density Structure of Spiral Arms}\label{sec:armdensity}

While there are numerous studies on the generation of stellar spiral
arms, they concentrate mostly on arm morphologies and longevity without
focusing on the arm density structures (e.g.,
\citealt{her90b,bar92,mih94,naa03}). Without much information on the
vertical structure of stellar spiral arms, most previous works that
studied galactic spiral shocks across the arms employed the arm
potentials that are independent of, or varying weakly with, the
vertical height (e.g., \citealt{kim06,kko06,kko10}). \cite{cox02}
suggested an analytic expression for 3D spiral density perturbations as
described below. In this section, we analyze the density structures of
spiral arms produced in our models and compare them with the analytic
suggestion of \cite{cox02}.

Starting from a physically-motivated trial function for the spiral
gravitational potential that drops off exponentially in the radial
direction, \cite{cox02} obtained a simple formula for the $m=2$ spiral
density perturbation
\begin{equation}\label{eq:rho_arm}
  \rho_1(R,\phi,z) = \rho_{0} \exp \left(- \frac{R-R_0}{R_s}\right)\sech^2\left(\frac{z}{h}\right)
  \sum_{n=1}^{N} a_n\cos(n\gamma),
\end{equation}
with
\begin{equation}\label{eq:gamma}
  \gamma = 2 \left[\phi - \phi_p(R_0)-\frac{\ln(R/R_0)}{\tan i}\right],
\end{equation}
where $\rho_0$ is the normalization coefficient, $R_{0}$ is the
fiducial radius, $R_{s}$ is the radial scale length of the perturbed
density, $h$ is the vertical scale height, $a_n$ is the Fourier
amplitude of an azimuthal mode $n$, and $\phi_p(R_0)$ denotes the
azimuthal phase of the arm at $R=R_0$. The summation over $n$ is to
allow for linear superpositions of various modes along the azimuthal
direction for two-armed, logarithmic spirals. As examples, \cite{cox02}
considered two different cases: \textit{sinusoidal} arms having $N=1$
with $a_1 = 1$ (e.g.\ \citealp{kim02,wad04,she06}) and
\textit{concentrated} arms having $N=3$ with $a_1 = 8/(3\pi)$, $a_2 =
1/2$, and $a_1 = 8/(15\pi)$. The concentrated arms have flatter
interarm regions than the sinusoidal arms. The four-armed concentrated
spirals were used by \citet{dob06} and \citet{dbp06} to study gas
dynamics driven by spiral arms. The thick spiral potential considered
by \citet{pat96} corresponds to the sinusoidal arms.

Figure~\ref{fig:rhoz_arm} plots the vertical density profiles in the
arm regions at $R=6\kpc$ and $\phi=90\deg$ of model  TA2H at $t=0.38$
when the arm strength is nearly peaked (left), and at $t=0.85$ when the
arms are in the decaying phase (right). The black solid, dotted, and
dot-dashed lines draw the total density $\rho(R,\phi,z)$, the
azimuthally-averaged density $\brho(R,z) = \int \rho(R,\phi,z) d\phi
/2\pi$, and the perturbed density $\rho_1 = \rho-\brho$, respectively.
The three density distributions are well described by
$\mathrm{sech}^2(z/h)$, shown as thin red lines, with $h=0.33$,
$\bar{h}=0.36$, $h_1=0.30\kpc$ at $t=0.38$ for the total,
azimuthally-averaged, perturbed densities, respectively, and with
$h=0.37$, $\bar{h}=0.36$, $h_1=0.41\kpc$ at $t=0.85$. The temporal
increase of the scale heights between $t=0.38$ and $0.85$ is due to
disk heating. But, these agree within 10\%, validating the vertical
dependence of $\rho_1$ in Equation \eqref{eq:rho_arm}, with a scale
height similar to that of the background averaged disk.

\begin{figure*}
\hspace{0.5cm}\includegraphics[angle=0, width=17cm]{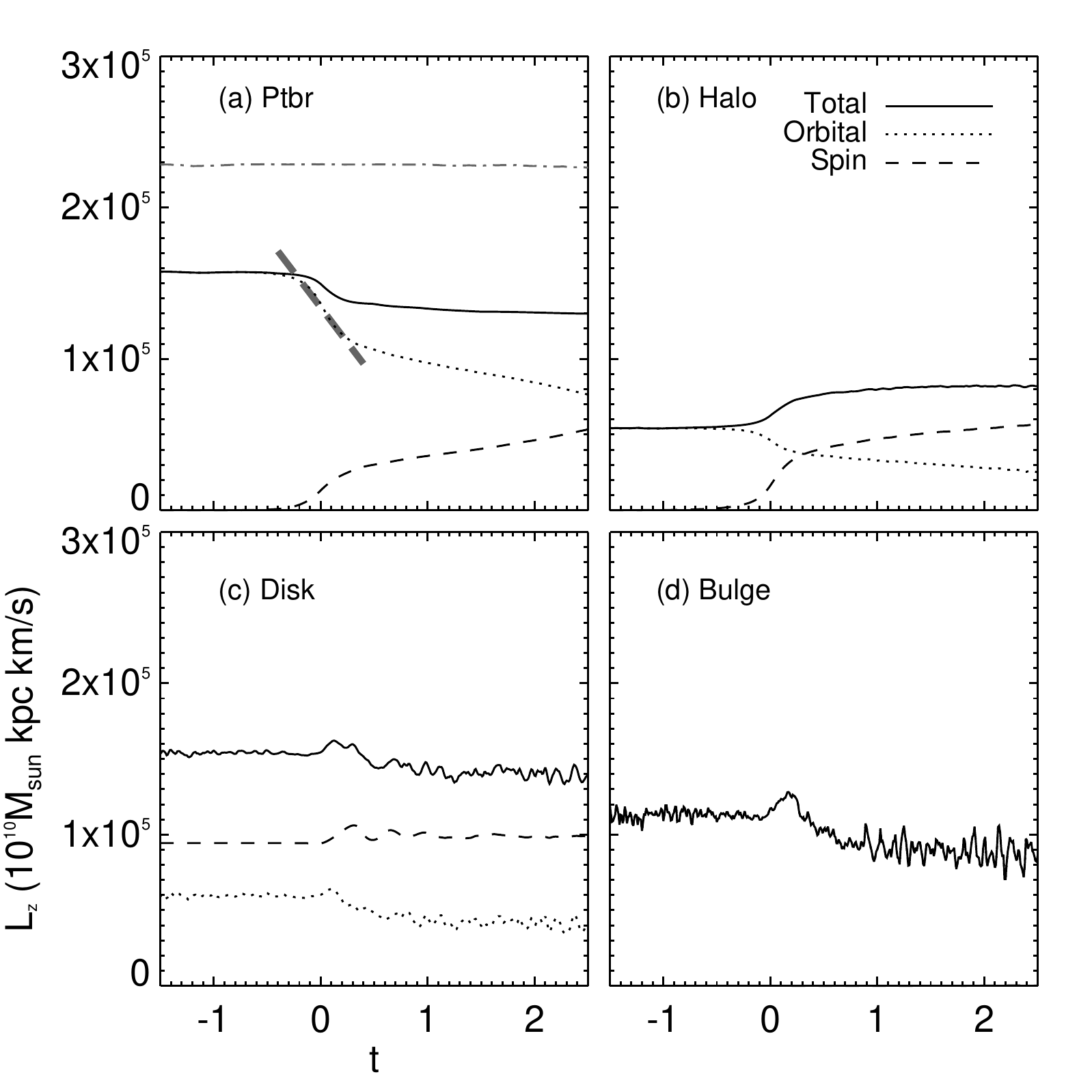}
\caption{\label{fig:angmom} Temporal variations of the net (solid),
orbital (dotted), and spin (dashed) angular momenta of (a) the
companion, and the (b) halo, (c) disk, and (d) bulge of the primary
galaxy during the tidal encounter of model TA2H. In (a), the dot-dashed
line plots the total angular momentum of the whole system, while the
thick dashed-line segment indicates the decrease of the orbital angular
momentum due to dynamical friction expected from Chandrasekhar formula.
The companion and the primary halo that do not rotate initially acquire
a significant amount of the spin angular momentum at the end of the
encounter. The orbital angular momentum of the disk/bulge experiences a
slight boost near $t=0$ and declines afterward.}
\end{figure*}

\begin{figure}
\hspace{0.1cm}\includegraphics[angle=0, width=8.5cm]{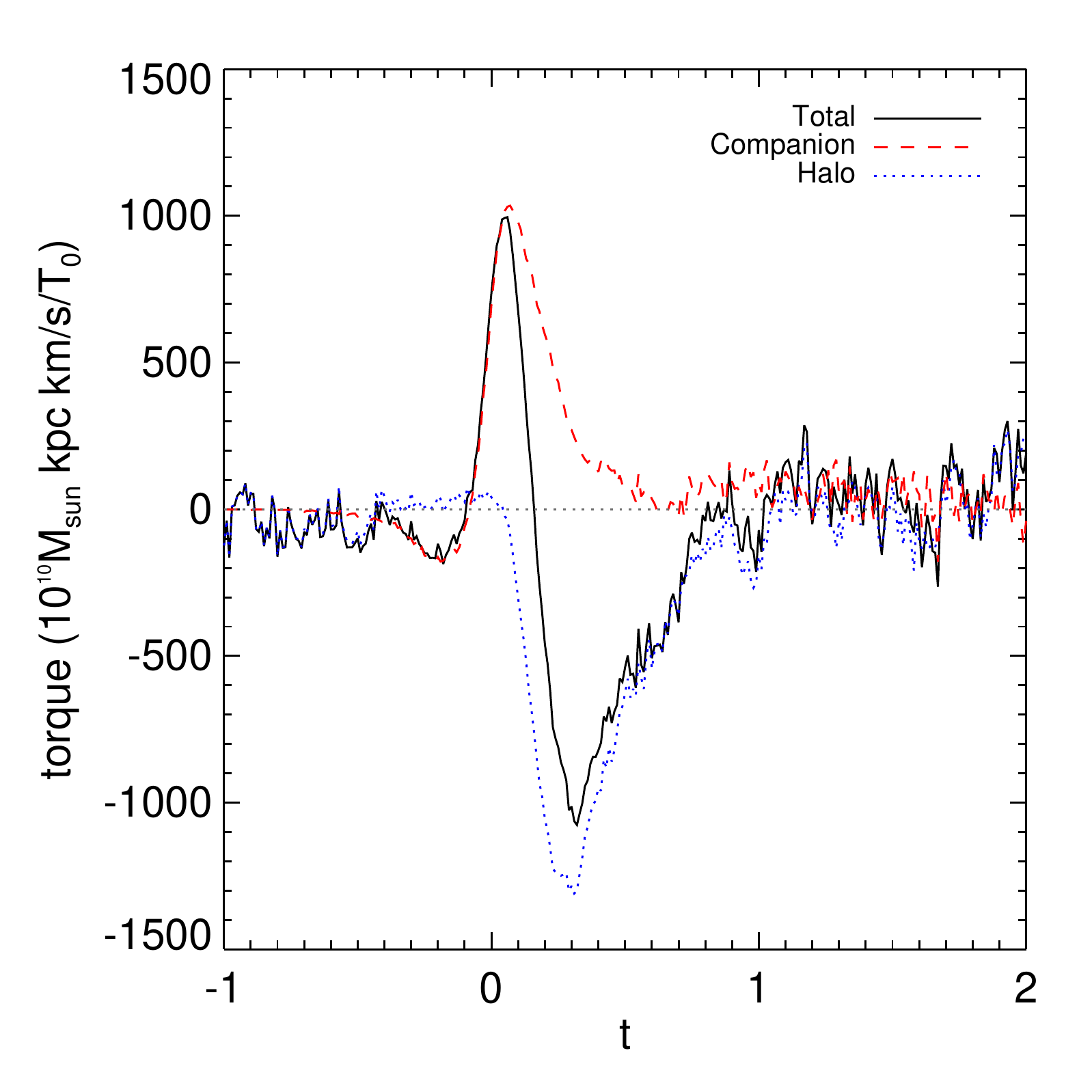}
  \caption{\label{fig:torque}
  Torques on the disk/bulge exerted by its own halo (dotted) and the companion
  (dashed) in model TA2H as functions of time. The solid line gives the
  total torque.  While the halo torque is always negative, the companion torque
  is positive over $-0.1\lesssim t \lesssim 0.6$, caused by the displacement of
  the disk/bulge from the center of mass of the halo (See
  Figure~\ref{fig:bulgedisp}). With the positive net torque, the
  disk/bulge gains angular momentum during $-0.1\lesssim t \lesssim
  0.15$. }
\end{figure}

\begin{figure*}
\hspace{0.5cm}\includegraphics[angle=0, width=17cm]{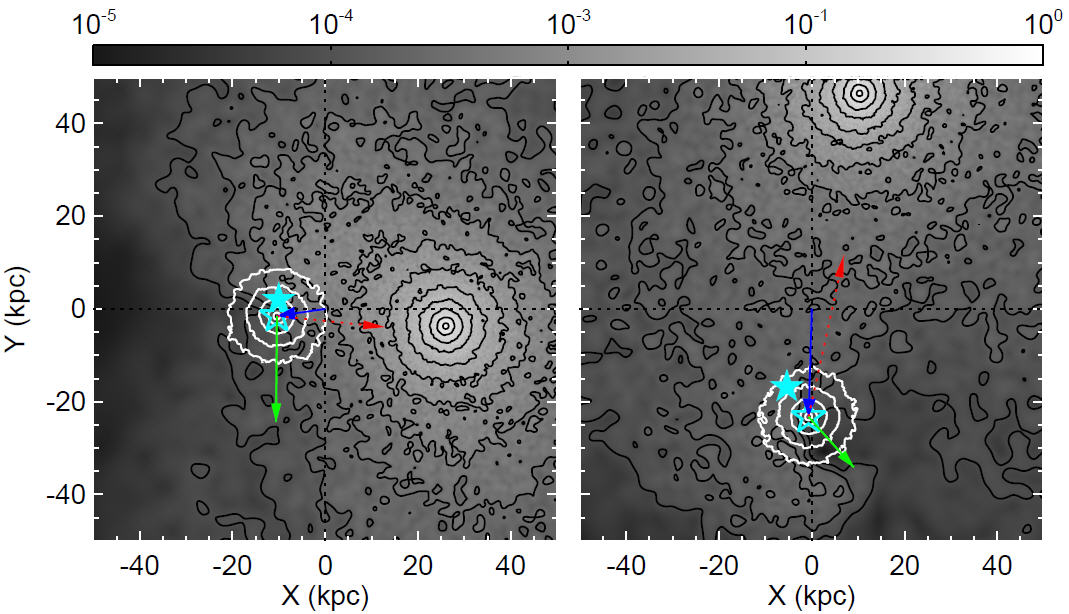}
  \caption{\label{fig:bulgedisp}
    Density distributions at the $z=0$ plane of the companion (grey scale and black contours) and the disk/bulge of the
    primary (white contours) of model TA2H at (left) $t=0$ and (right)
    $t=0.25$. The filled and open star symbols in cyan mark the centers of mass of the
    halo and disk/bulge of the primary, respectively. The blue and
    green arrows in each panel are the position and velocity vectors of the
    disk/bulge in the center of mass frame of the whole system, respectively, while
    the red dotted arrow indicates the direction from the disk/bulge to the
    companion. The greyscale bar labels $\log(\rho/1 \dunit)$.  The contours mark $\log(\rho/1 \dunit) = -1.0, -1.5, -2.0,
    \cdots$ from inside to outside.}
\end{figure*}

Figure~\ref{fig:rhor_arm} plots as solid lines $\rho_1$ as a function
of $R$ along the cuts with $\phi= \pm 90\deg$ for model TA2H in the
midplane. The top and second rows are at $t=0.38$, while the third and
bottom rows are at $t=0.85$. The thin solid lines give the simulation
results, while the thick grey lines draw the best fits using Equation
\eqref{eq:rho_arm} based on the sinusoidal arms (left panels) and the
concentrated arms (right panels).  The fact that Equation
\eqref{eq:rho_arm} describes the arm positions represented by the
over-densities fairly well implies that the spiral arms are closely
logarithmic in shape. Overall, the concentrated arms fit the radial arm
structure better at $t=0.38$ when the arms are strong with $\Farm
\gtrsim 10\%$, whereas the sinusoidal arms provide a better fit at
$t=0.85$ when the arms are weak with $\Farm \lesssim 10\%$.  We note
however that these fits are reasonably good only at $R\gtrsim 4 \kpc$
and fail at smaller $R$. This is expected from the fact that the
tidally-driven arms do not decline monotonically with $R$ but are
peaked at $ 5\kpc <R<10\kpc$ (see Fig.~\ref{fig:F-R}), while Equation
\eqref{eq:rho_arm} requires arms to decay exponentially in the radial
direction over all radii.

\begin{figure*}
\hspace{0.cm}\includegraphics[angle=0, width=17cm]{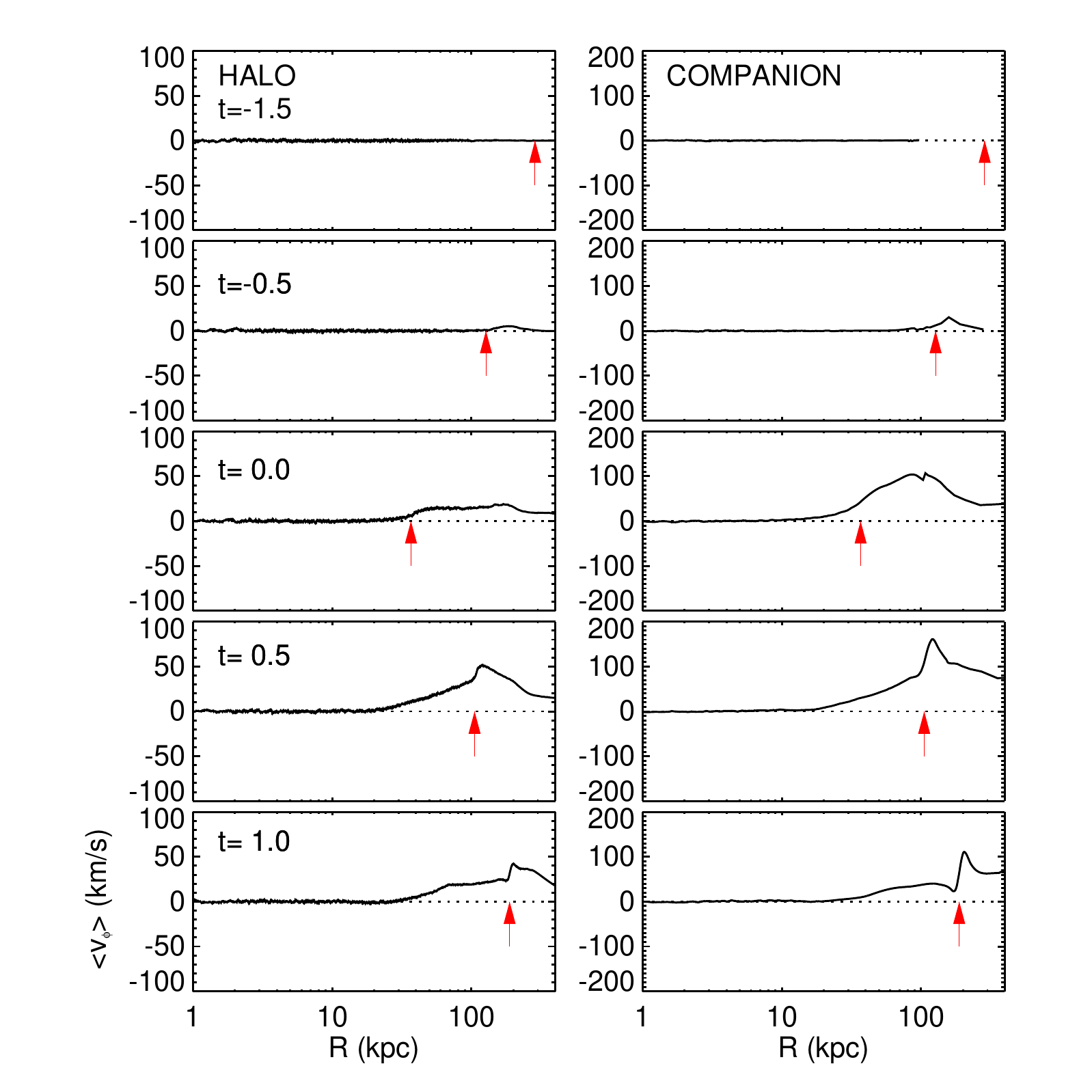}
   \caption{\label{fig:vtiso_halo}
     Axially-averaged rotation velocities $\langle v_\phi\rangle$ of the
     primary halo (left) and the companion (right) as functions of the
     radial distance in model TA2H. The vertical arrow in each panel
     marks the location of the companion in the left panels and the primary halo
     in the right panels.}
\end{figure*}

\section{Orbital Decay}\label{sec:orbit_decay}

As Figure~\ref{fig:snapshots} shows, the actual trajectories of the
primary and companion galaxies during the tidal encounter deviate from
those under the assumption of the rigid halos. The discrepancy is due
to dynamical friction occurring when two galaxies overlap at least
partly. As the companion enters the primary halo, the former creates a
density wake in the latter that in turn exerts gravitational drag force
on the former. The drag force removes the orbital angular momentum from
the companion, decreasing the pericenter distance. The orbital angular
momentum extracted from the companion is transferred to the spin
angular momentum of the primary halo. A similar process also occurs
when the primary enters the companion halo, which results in the
angular momentum transfer from the orbit of the primary to the spin of
the companion.

To quantify the amount of the angular momentum transfer,
Figure~\ref{fig:angmom} plots the temporal variations of the
$z$-component of the angular momenta of all the components as functions
of time. In each panel, the solid line draws the net (orbital plus
spin) angular momentum, while the dotted and dashed lines give the
orbital and spin angular momenta defined, respectively, by
 \begin{equation}
 \mathbf{L}_{\rm orb} = M \mathbf{X}_{\rm CM}\times \mathbf{V}_{\rm CM},
 \end{equation}
and
 \begin{equation}
  \mathbf{L}_{\rm spin} = \sum m_i (\mathbf{x}_i -
 \mathbf{X}_{\rm CM})\times (\mathbf{v}_i - \mathbf{V}_{\rm CM}),
 \end{equation}
where $M=\sum m_i$ is the total mass, $\mathbf{X}_{\rm CM} = \sum m_i
\mathbf{x}_i / M$ is the position vector of the center of mass, and
$\mathbf{V}_{\rm CM} = \sum m_i \mathbf{v}_i / M$ is the mean velocity
of each component, measured in the center-of-mass frame of the whole
system.  The dot-dashed line in Figure~\ref{fig:angmom}(a) plots the
total angular momentum of the whole system, showing that it is
conserved within 3\% throughout the entire evolution. Initially, all
the components except the disk do not spin, and their angular momenta
are dominated by the orbital motions. As the primary and companion
approach the pericenter, they start to experience dynamical friction,
and the primary halo and the companion begin to continuously acquire
the spin angular momentum at the expense of  the orbital angular
momentum.

In order to check if the decrease in the orbital angular momentum of
the companion is really caused by dynamical friction due to the primary
halo, we use the \citet{cha43} formula for the drag force
\begin{eqnarray}\label{eq:df_chandra}
  \mathbf{F}_{\rm DF} =
     - \frac{4 \pi G^2 \Mcomp^2 \bar\rho_h \ln \Lambda}{v^3_{\rm ptb}}
  \left[\mathrm{erf}(X) - \frac{2X}{\sqrt{\pi}}e^{-X^2}\right]
  \mathrm{\mathbf{v}}_{\rm ptb},
\end{eqnarray}
where $\bar\rho_h$ is the average density of the halo, $v_{\rm ptb}$ is
the relative speed of the companion, $X = v_{\rm ptb} /
(\sqrt{2}\sigma)$, with $\sigma$ denoting the mean velocity dispersion
of halo particles, and $\ln \Lambda = \ln (b_{\rm max}/b_{\rm min})$ is
the Coulomb logarithm introduced to avoid a singularity in the force
evaluation, with $b_{\rm max}$ and $b_{\rm min}$ representing the
maximum and minimum impact parameters of the background particles,
respectively. For orbital decay of a satellite galaxy in a dark matter
halo, Equation (\ref{eq:df_chandra}) matches the numerical results
provided $b_{\rm max}$ is taken to be of order of the orbital radius of
the halo, and $b_{\rm min} \approx \max(G\Mcomp/v^2_{\rm ptb}, r_h)$
with $r_h$ being the half-mass radius (e.g., \citealt{lin83,wei86}). To
apply Equation (\ref{eq:df_chandra}) to model TA2H at $t=0$, we
calculate the mean density of the halo as $\bar\rho_h= \int
\rho_h(|\mathbf{r}-\mathbf{r}_h|) \rho_{\rm
ptb}(|\mathbf{r}-\mathbf{r}_{\rm ptb}|) d^3 r / \int \rho_{\rm ptb}
(|\mathbf{r}-\mathbf{r}_{\rm ptb}|) d^3 r = 3.0\times 10^{5}
\Msun\kpc^{-3}$, where $\mathbf{r}_h$ and $\mathbf{r}_{\rm ptb}$ denote
the position vectors of the halo and the companion, respectively. The
velocity of the primary relative to the halo is $v_{\rm ptb} =
380\kms$. With $b_{\rm max} = \Rperi$ and $b_{\rm min}=
(1+\sqrt{2})r_{\rm ptb}$, $\ln \Lambda = 1.3$. The resulting torque on
the companion is given by
 \begin{equation}
 \frac{d\mathbf{L}_{\rm orb, ptb}}{dt} = -\mathbf{R}_{\rm peri}\times \mathbf{F}_{\rm DF},
 \end{equation}
which is plotted as a thick solid line in Figure~\ref{fig:angmom}(a),
in fairly good agreement with the instantaneous decreasing rate of the
orbital angular momentum of the companion.  This validates that
dynamical friction is indeed a primary cause of the angular momentum
transfer in our numerical simulations.

A close inspection of Figure~\ref{fig:angmom} reveals that while the
orbital angular momenta of the halo and companion decrease
monotonically with time, those of the disk and bulge stay almost
constant at $t<-0.1$, increase slightly for $-0.1 \lesssim t \lesssim
0.15$, and subsequently decrease slowly. This indicates that the disk
and bulge experience positive torque temporarily near the pericenter,
while the halo and companion always receive negative torque. To analyze
what causes this unexpected behavior of the angular momenta of the disk
and bulge, Figure~\ref{fig:torque} plots as dashed and dotted lines the
torques exerted on the disk/bulge due to particles in the companion and
the halo, respectively. It is apparent that the torque by the companion
is overall positive, while the halo gives a negative torque to the
disk/bulge. The resultant net torque, shown as the solid line, is
positive for $-0.1 \lesssim t \lesssim 0.15$, which boosts the orbital
motion of the disk/bulge system.

To understand why the companion torque is positive,
Figure~\ref{fig:bulgedisp} plots the density distributions of the
companion (gray scale with black contours) and the disk and bulge of
the primary (white contours) at $t=0$ (left) and $t=0.25$ (right) of
model TA2H in the orbital plane. The contour levels decrease by a
factor of $10^{0.5}$ starting from the innermost level of $0.1 \dunit$
for both companion and disk/bulge. Note a clear density wake in the
companion formed at $t=0$ near $(X,Y)\sim(0,20)\kpc$ due to the
primary. The filled and open star symbols in cyan mark the centers of
mass of the halo and disk/bulge of the primary, respectively, while the
arrows in green indicate the velocity vectors of the disk/bulge in the
center of mass frame of the whole system.

The dynamical friction of the primary due to the companion is stronger
for the (outer) halo than the disk/bulge owing to proximity to the
companion. This causes the halo to move slower than the rest of the
primary galaxy, gradually making them displaced from each other. At
$t=0$, the center of mass of the disk/bulge is entering the third
quadrant and moving in the negative-$Y$ direction, while that of the
halo is still in the second quadrant. At this time, the position vector
of the bulge, indicated by the blue arrow, is inclined to the line
connecting the centers of mass of the disk/bulge and the companion (the
red dotted arrow). In such a geometrical configuration, the companion
pulls the disk/bulge forward by providing a positive torque, while the
halo tends to slow them down. It turns out that the positive torque is
stronger than the gravitation pull by the halo at this time, as
Figure~\ref{fig:torque} shows. The net effect is that the separation
between the centers of mass of the disk/bulge and the halo increases at
a faster rate. The positive torque  becomes strongest $t=0.06$ when the
companion is closest to the disk/bulge, and then decreases as the
companion moves away. At $t=0.25$, the positive torque by the companion
is smaller than the negative torque by the halo. The disk/bulge is
subsequently pulled back to, and moves in phase with, the halo after
$t\sim 1$.

Does the dynamical friction between two galaxies really transfer the
orbital angular momentum of one galaxy to the spin angular momentum of
the other? Figure~\ref{fig:vtiso_halo} plots the temporal changes of
the axially-averaged rotation velocities $\langle v_{\phi}\rangle$ of
the halo (left) and companion (right). The vertical arrow in each panel
marks the position of the companion and the primary in the left and
right panels, respectively. Initially ($t=-1.5$), they are
non-rotating, spherical, and about $280\kpc$ apart. As the center of
the companion (halo) moves through the outer parts of the primary halo
(companion), it creates a density wake in the halo (companion) that
pulls it backward. At the same time, the wake in the halo (companion)
is pulled \emph{forward} by the companion (halo), acquiring positive
rotational velocities as shown in Figure~\ref{fig:vtiso_halo}. The
mass-weighted mean rotational velocities of the companion and the halo
at $t = 1.0$ is $\langle v_{\phi} \rangle \sim 15 \kms$ and $7 \kms$,
respectively. This suggests that tidal interactions can be a source of
the spin angular momentum of a dark matter halo that might be
non-rotating when it first formed.

\section{SUMMARY \& DISCUSSION}\label{sec:summary}

\subsection{Summary}

Using self-consistent, 3D $N$-body simulations, we have investigated
the physical properties of tidal features induced in a  disk galaxy
interacting with its companion.  The disk galaxy consists of a halo, a
bulge, and a stellar disk, corresponding to Milky-Way type galaxies,
all of which are represented by live particles. The perturbing
companion is simplified by a live halo alone. This work is a
straightforward extension of Paper~I that considered a 2D, razor-thin
disk residing within a fixed halo. We do not consider the effect of the
gaseous component in the present work. By varying the companion mass
and the pericenter distance, we explore the intermediate and weak
interaction regimes with the strength parameter of $0.04 \lesssim P
\lesssim 0.41$ or $ 0.06 \lesssim S \lesssim 0.25$ (see Eqs.\
[\ref{eq:P}] and [\ref{eq:S}] for the definitions of $P$ and $S$). We
analyze the properties of tidal tails formed in outer regions, and
spiral arms in the inner regions, and study their dependence on $P$ and
$S$. We also study the 3D density structures of spiral arms as well as
the orbital decay of the galaxies caused by dynamical friction
occurring during a tidal encounter. The main results of this work can
be summarized as follows.

\begin{enumerate}
\item
  The tidal force from the companion excites epicycle motions in the
  outer regions of the primary, forming a bridge at the near side and an
  extended tail at the far side for interactions with $P\gtrsim
  0.05$ or $S\gtrsim 0.07$: tidal forcing weaker than this is unable to
  produce a tidal tail. The tail formation time, pitch angle, and surface
  density measured at $R=20\kpc$ scale as
  $t_{\mathrm{tail}} = 0.07  P^{-0.49}$,
  $\tan i_{\mathrm{tail}} = 0.1 P^{0.24}$, and
  $\Sigma_{\mathrm{tail}}/\Sigma_{20} = 60 P^{0.65}$ in terms of $P$,
  and $t_{\mathrm{tail}} = 0.04  S^{-0.79}$,
  $\tan i_{\mathrm{tail}} = 1.25 S^{0.43}$, and
  $\Sigma_{\mathrm{tail}}/\Sigma_{20} = 144 S^{1.1}$ in terms of $S$.
  This is in good agreement with the 2D results of Paper~I, although
  the 2D tails are slightly stronger than the
  3D counterparts, due to stronger gravity in the disk
  midplane. We also find that the tail response $\Ttail$ defined by the
  fraction of the disk particles consisting of a tail at $R > 10R_d$ has
  an excellent correlation with $S$ as $\Ttail = 0.31 S^{2.1}$.

\item
  The tidal force also excites two-armed spiral density waves over
  a range of radii in the inner parts of the disk.  We calculate the spatial and temporal
  dependence of the arm strength $\Farm$ defined by the ratio of the
  maximum gravitational force due to the $m=2$ modes to the centrifugal force
  of disk rotation (Eq.~[\ref{eq:F}]).
  For intermediate tidal interactions with $P>0.17$ (or $S>0.13$),
  the arms are strongest at $R\sim5$--$10\kpc$, while weak interaction
  models with $P<0.1$ (or $S<0.12$) have the arms induced at $R\sim 8$--$15\kpc$.
  The spiral arms are stronger in models
  with larger $P$ or $S$, with the peak arm strength given by
  $\Fmax=0.39P^{0.7}$ or
  $\Fmax=0.82S^{1.1}$. The spiral arms are approximately logarithmic in shape,
  with a pitch angle $i\sim 15\deg-20\deg$ when the arms are strongest.
  After the peak, the arms decay with a timescale of $\sim0.5\Gyr$, and
  wind out as $\tan i \propto t^{-0.75}$.
  The derived arm pattern speed $\Omega_p$ is a decreasing function of $R$
  and becomes close to the $\Omega-\kappa/2$ curve at late time.
  This suggests that the induced
  spiral arms are kinematic density waves modified weakly by
  self-gravity.
  Compared to the 2D counterparts, arms in 3D models are weaker,
  have a smaller pitch angle, and wind and decay
  more rapidly with time, due to weaker gravity.

\item
  When compared with the analytic expression of \cite{cox02},
  the 3D density structure of tidally-induced, $m=2$ arms can
  be reasonably well described by the concentrated arms when the arms are in the
  nonlinear regime with  $\Farm \gtrsim 10\%$. On the other hand,
  arms in the linear regime with $\Farm \lesssim 10\%$ are better fitted by
  the sinusoidal model.
  The perturbed density in the arm regions
  follows the characteristic profile $\rho_1(z) \propto \mathrm{sech}^2 (z/h_1)$
  along the vertical direction,
  with the scale height $h_1$ not much different from that of
  the initial disk.

\item
  The halos of the primary and companion
  become partially overlap in the course of a tidal encounter. This creates
  gravitational density wakes in the halos, resulting in dynamical
  friction. The dynamical friction has a few notable effects. First,
  it transports the orbital angular
  momentum of one galaxy to the spin angular momentum of the halo of its companion,
  making the pericenter distance smaller than that under the rigid halos.
  Second, the initially rotation-free halos achieve the mass-weighted
  rotational velocities amounting to $\langle v_\phi \rangle \sim 15\kms$ and $7\kms$ for
  the primary and companion, respectively. Third, the dynamical
  friction is stronger for the halo, causing offsets between the
  centers of mass of the halo and disk/bulge of the primary. Thus,
  the disk/bulge can temporarily gain angular momentum by a positive torque
  from the companion near the pericenter,
  although they are later pulled back to the center of mass of the halo as
  the companion moves away.

\end{enumerate}

\subsection{Discussion} \label{sec:discussion}

Tidal interaction of a galaxy with its companion in a prograde orbit is
studied using fully consistent $N$-body models. We find that the
physical properties of the tidally-induced nonaxisymmetric features are
tightly correlated with the interaction strength parameters $P$ and
$S$. These correlations are in qualitative agreement with those from
the 2D results presented in Paper~I, although the structures in the 3D
models are generally weaker and decay more rapidly. These are also
qualitatively consistent with the results of \cite{elm91} who described
in terms of $S$ the formation criteria of the outer disk features such
as double arms corresponding to spiral arms decoupling from the tidal
tail. \cite{byr92} similarly found that the tidal strength $P$
determines the strength of grand-design spiral arms in interacting
galaxies. Overall, $S$ is a better parameter for characterizing tidal
tails formed in the outer regions, while spiral arms induced in the
inner regions are better correlated with $P$.

Strength and shapes of tidal tails have been used to constrain the
masses of dark halos (e.g., \citealt{fab79}). Using numerical
simulations, \citet{dub96} found that a tidal tail becomes shorter and
less massive as the mass and size of a dark halo increase due to a
shorter interaction time as well as a deeper potential well (e.g.,
\citealt{whi82}). More quantitatively, \cite{mmw98} suggested $\eps
\equiv (v_{e}/v_{c})^2$ as an indicator of the susceptibility of a disk
galaxy to the tail formation, where $v_e$ and $v_c$ refer to the escape
and circular velocities at $R=2R_d$, respectively. Note that $\eps$
measures the depth of the halo potential relative to the specific
kinetic energy of the disk. Using galaxy models with $4 < \eps< 8$,
\citet{spr99} showed that the maximum tidal response $\Teff$ described
in Section \ref{sec:tail} is inversely proportional to $\eps$. They
argued that tidal forcing produces a tail with a substantial amplitude
only when $\eps \leq 6.5$ in their models. \citet{dub99} conducted an
extensive survey of galaxy collisions, and confirmed that $\eps$ is
really a governing parameter and must be less than $6.25$ for the tail
formation.

Our primary galaxy has $\eps =4$, corresponding to the lower end in the
models of \citet{spr99} (see their Figure 10). Based on the results of
\citet{spr99} and \citet{dub99}, therefore, one can naturally expect
strong tidal tails produced in all of our simulations. However, it
turns out that a tail is absent in model TB3 with $S=0.07$, while all
the other models with $S>0.09$ do produce a tail with strength
depending on $S$. Model TB3 has a very weak tidal response at
$\Teff=0.005$ from particles captured by the companion, even if it is
more prone, in terms of $\eps$, to the tail formation than any model of
\citet{spr99}. This suggests that the tail formation should depend not
only on $\eps$ but also on the interaction strength. Our numerical
results imply that a criterion for producing a tidal tail is $S>0.09$
(or  $P>0.05$) for $\eps=4$. The critical values of $S$ and $P$ would
increase with increasing $\eps$, since it would then become
increasingly difficult to make disk particles climb out of the halo
potential well.

We found that spiral arms induced by a tidal encounter are
approximately logarithmic, consistent with many of observed
grand-design spirals including M51 (e.g.,
\citealp{ken81,elm89,pat06,she07}). In our models, a logarithmic shape
results from the fact that the spiral arms in the present simulations
are close to kinematic density waves, slightly modified by
self-gravity, for which the pitch angle varies with time and space as
 \begin{equation}
 \tan i = t^{-1} \left| \frac{d(\Omega - \kappa/2)}{d\ln R}
 \right|^{-1}
 \end{equation}
(e.g., Eq.~(6.26) of \citealt{bt08}). The rotation curve of our galaxy
model shown in Figure \ref{fig:rotprofile} gives $d (\Omega -
\kappa/2)/d\ln R\sim 3.5\pm0.5\kmskpc$ at $5\kpc \le R \le15\kpc$ where
arms are induced, making $\tan i$ roughly constant over $R$. This
predicts that kinematic density waves wind out with time as $\tan
i\propto t^{-1}$ when self-gravity is neglected. Since $\Omega >
\kappa/2$ for galactic disks, this also predicts that $\tan i$ is
smaller for larger $|d\Omega/d\ln R|$, consistent with the
observational results that the arm pitch angle has a negative
correlation with the shear rate (e.g., \citealt{sei05,sei06,sei14}).
Our results show that self-gravity reduces the arm-winding rate to
$\tan i\propto t^{-0.75}$, although it is not strong enough to maintain
the arms quasi-steady. This holds even in 2D models with overestimated
self-gravity at the disk midplane (Paper~I).

Another prediction of kinetic density waves is that the arm pattern
speed is not strictly constant over the radius but a decreasing
function of $R$.  Indeed, \cite{wes98} found that the pattern speed of
the arms in M81 interacting with NGC 3077 is best described by
$\Omega_p= 44.6$--$2.3 (R/1\kpc)\kmskpc$ based on the \citet{tre84}
method, after taking allowance for the radial variation of $\Omega_p$.
\citet{mei13} applied a similar method to M51 in obvious interaction
with NGC 5195, and found that the arms consist of multiple patterns
with larger pattern speeds at smaller radii (see also
\citealt{mei08a,mei08b}), although it is questionable whether they are
really multiple patterns with different pattern speeds or a single
pattern with radially-varying $\Omega_p$. Realistic interaction models
of \citet{sal00b} for the M51/NGC 5195 pair also predict $\Omega_p$
close to the $\Omega-\kappa/2$ curve over a range of radii where the
arms are strong. It is interesting to note the spiral arms in the
isolated galaxy NGC 1068 have a pattern speed that decreases steeply
with $R$, indicative of a short lifetime of order of $\sim0.1\Gyr$
(\citealt{spe11}; see also \citealt{spe12}). All of these suggest that
arms in real spiral galaxies are more likely transient rather than
possessing the characteristics of quasi-stationary density waves unless
the disks are strongly self-gravitating.

While we employ very simple galaxy models for tidal interactions, our
models can well be applied to the M51/NGC 5195 system that have the
mass ratio of $\sim0.3$--$0.55$ (e.g., \citealt{smi90}) and a
pericenter distance of $\sim20$--$30\kpc$ (e.g., \citealt{sal00a}),
corresponding to models TA1 and TA2. As Figure \ref{fig:F-R} shows, the
arm strength is maximized $\sim200\Myr$ after the pericenter passage.
\citet{lee05} found observationally that the age distribution of star
clusters in the arms of M51 has a narrow peak at $4$--$10\Myr$ and a
broad peak at $100$--$400\Myr$. If the enhanced star formation is
really triggered by tidal effects, the M51/NGC 5195 pair might have
undergone double interactions, the first one $400$--$500\Myr$ ago and
the second one $50$--$100\Myr$ ago \citep{sal00a}, considering the time
delay between the arm formation and the closest approach.  The peak arm
strength in models TA1 and TA2 is $\Farm\sim10-20\%$, consistent with
the radially averaged value of $\sim15-20\%$ from $K$-band observations
(e.g., \citealt{rix93,rix95,sal00b,sco01}). The arm pitch angle at peak
is $i\sim 17\degr-22\degr$ in models TA1 and TA2 (Figure
\ref{fig:winding}), which is also  consistent with the observed pitch
angle of $i\sim 17.5\degr-21.1\degr$ (e.g.,
\citealt{she07,fle11,hu13,pue14}).  Based on our numerical results, the
stellar spiral arms in M51 are in the nonlinear regime and their 3D
density structure may be described by the concentrated arm model of
\citet{cox02}.

By analyzing the properties of $m=2$ spiral arms in the \emph{Spitzer}
Infrared Nearby Galaxies Survey, \citet{ken15} very recently found that
spiral morphology depends only weakly on the galaxy properties, while
the arm strength is tightly correlated with tidal forcing from nearby
companions. Their Figure 17 shows that $\mathcal{R} \equiv
\tilde\Sigma_{m=2}/\tilde\Sigma_{m=0}$ is an increasing function of
$P$, which agrees with our numerical results qualitatively, but
seemingly not quantitatively. That is, $\mathcal{R}$ of the arms in the
\emph{Spitzer} samples varies in the range of $0.05$--$0.5$ for $\log
P=-5$--$0$, while our results predict $\mathcal{R} \sim 0.2$--$1$ only
for $\log P\simgt -2$ and no arm induced for $\log P \simlt -2$ (see
also \citealt{byr92}). As \citet{ken15} noted, this apparent
quantitative discrepancy may arise because the observed $P$ values in
the \emph{Spitzer} samples correspond to the projected distance of a
nearest neighbor at the current epoch, while $P$ in our models measures
the tidal force using the 3D distance at the epoch of the pericenter
passage.  The projection effect overestimates the tidal forcing, while
the distance at the current epoch is likely to underestimate the true
$P$ significantly. For example, a comparison between Figure 17 of
\citet{ken15} and our Figure \ref{fig:m2m0_P} suggests that NGC 1566
with $\mathcal{R} \sim 0.2$ and the current forcing estimate of $\log
P\sim -4.3$ might probably have undergone an interaction with its
companion NGC 1581 with $\log P\sim -1$ at the pericenter passage.

Finally, we remark on the absence of a bar in our numerical
simulations.  As Figures \ref{fig:surf_xy} and \ref{fig:surf_phir}
show, tidal forcing in our models induces spiral structures at $R\simgt
5\kpc$, while the inner regions with $R\simlt 4\kpc$ remain almost
unaffected.  This is in contrast to the results of \citet{nog87},
\citet{ger90}, and \citet{mih97} who showed that disks subject to tidal
interaction can be unstable to form a bar. The major difference between
the galaxy models used by the present paper and their work is that our
models possess a relatively strong bulge, while their models have no or
very weak bulge. The strong bulge in our models puts an inner Lindbland
resonance which suppresses feedback to swing amplification that would
otherwise produce a bar near the center \citep{too81}.  This suggests
that a strong bulge may be responsible for the absence of a bar in
non-barred grand design spiral galaxies.

\acknowledgments

We are grateful to the referee for a thoughtful and constructive
report. This work was supported by the National Research Foundation of
Korea (NRF) grant, No.~2008-0060544, funded by the Korea government
(MSIP). The computation of this work was supported by the
Supercomputing Center/Korea Institute of Science and Technology
Information with supercomputing resources including technical support
(KSC-2014-C3-003).

\end{document}